\documentclass[12pt]{article}

\title{Experimental Characterization of the Passive Mechanical Behavior of Pregnant Rat Pelvic Floor Muscles}

\usepackage{authblk}
\usepackage[numbers]{natbib}

\author[1]{Jose L. Monclova}
\author[2]{Matthew Shankel}
\author[1]{Ava M. English}
\author[2]{Mona Eskandari}
\author[1]{Heidi P. Feigenbaum}

\affil[1]{Department of Mechanical Engineering, Northern Arizona University, Flagstaff, Arizona, USA}
\affil[2]{Department of Mechanical Engineering, University of California Riverside, Riverside, California, USA}

\date{July 2026}

\usepackage{setspace} 
\usepackage[margin=1in]{geometry}
 \usepackage{csquotes}

\usepackage[acronym,
    style=list,
    nonumberlist,
    shortcuts,
]{glossaries}
\makeglossaries
\usepackage{relsize}

\usepackage[figuresleft]{rotating}
\usepackage{graphicx}
\usepackage{epstopdf}
\usepackage{placeins}
\usepackage{pgffor}
\usepackage{caption}
\usepackage{subcaption}
\usepackage{alphalph}
\usepackage{xcolor}
\usepackage{etoolbox}
\usepackage{pdflscape}
\usepackage{makecell}
\usepackage{longtable}
\usepackage{booktabs}
\usepackage{array}
\usepackage{lineno}
\usepackage{hyperref}
\usepackage{cleveref}

\hypersetup{
    colorlinks=true,
    citecolor=black,
    linkcolor=black,
    urlcolor=blue,
}

\newcommand{\RawSlopeSubfigure}[1]{%
    \begin{subfigure}[t]{\textwidth}
        \centering
        \ifnum#1<10
            \includegraphics[width=0.90\textwidth]
            {Ch99_SupplementaryInfo/Supp_Figures/Supp_RawSlopeCurves/FigS0#1_Savitsky_Golay.pdf}
        \else
            \includegraphics[width=0.90\textwidth]
            {Ch99_SupplementaryInfo/Supp_Figures/Supp_RawSlopeCurves/FigS#1_Savitsky_Golay.pdf}
        \fi
        \subcaption{\getdatasetname{#1}}
        \label{fig:raw-slope-#1}
    \end{subfigure}%
}

\newcommand{\getdatasetname}[1]{%
    \ifcase#1%
        \or 
        \or 
        \or Control frozen rat 2, with 10 cycles of 10\% strain amplitude preconditioning at 1\% strain per second
        \or Pregnant frozen rat 3, with 5 cycles of 10\% strain amplitude preconditioning at 1\% strain per second
        \or Pregnant frozen rat 3, with 10 cycles of 10\% strain amplitude preconditioning at 1\% strain per second
        \or Control frozen rat 4, with 5 cycles of 10\% strain amplitude preconditioning at 1\% strain per second
        \or Control frozen rat 4, with 10 cycles of 10\% strain amplitude preconditioning at 1\% strain per second
        \or Pregnant frozen rat 5, with 5 cycles of 10\% strain amplitude preconditioning at 1\% strain per second
        \or Pregnant frozen rat 5, with 10 cycles of 10\% strain amplitude preconditioning at 1\% strain per second
        \or Control fresh rat 7, with 10 cycles of 10\% strain amplitude preconditioning at 1\% strain per second
        \or Control fresh rat 7, with 0 cycles of 10\% strain amplitude preconditioning at 1\% strain per second
        \or Control fresh rat 8, with 5 cycles of 10\% strain amplitude preconditioning at 1\% strain per second
        \or Pregnant fresh rat 9, with 10 cycles of 10\% strain amplitude preconditioning at 1\% strain per second
        \or Pregnant fresh rat 9, with 0 cycles of 10\% strain amplitude preconditioning at 1\% strain per second
        \or Pregnant fresh rat 10, with 0 cycles of 10\% strain amplitude preconditioning at 1\% strain per second
        \or Pregnant fresh rat 10, with 5 cycles of 10\% strain amplitude preconditioning at 1\% strain per second
        \or Pregnant fresh rat 12, with 2 cycles of 10\% strain amplitude preconditioning at 0.5\% strain per second
        \or Control fresh rat 13, with 10 cycles of 10\% strain amplitude preconditioning at 0.5\% strain per second
        \or Pregnant fresh rat 14, with 10 cycles of 10\% strain amplitude preconditioning at 0.5\% strain per second
        \or Pregnant fresh rat 14, with 2 cycles of 10\% strain amplitude preconditioning at 0.5\% strain per second
        \or Control fresh rat 15, with 2 cycles of 10\% strain amplitude preconditioning at 0.5\% strain per second
        \or Control fresh rat 15, with 0 cycles of 10\% strain amplitude preconditioning at 0.5\% strain per second
        \or Pregnant fresh rat 16, with 0 cycles of 10\% strain amplitude preconditioning at 0.5\% strain per second
        \or Pregnant fresh rat 16, with 2 cycles of 10\% strain amplitude preconditioning at 0.5\% strain per second
        \or Control fresh rat 17, with 15 cycles of 10\% strain amplitude preconditioning at 0.5\% strain per second
        \or Control fresh rat 18, with 0 cycles of 10\% strain amplitude preconditioning at 0.5\% strain per second
        \or Control fresh rat 18, with 5 cycles of 10\% strain amplitude preconditioning at 0.5\% strain per second
        \or Pregnant fresh rat 19, with 5 cycles of 10\% strain amplitude preconditioning at 0.5\% strain per second
        \or Pregnant fresh rat 19, with 15 cycles of 10\% strain amplitude preconditioning at 0.5\% strain per second
        \or Pregnant fresh rat 20, with 15 cycles of 10\% strain amplitude preconditioning at 0.5\% strain per second
        \or Control fresh rat 21, with 10 cycles of 10\% strain amplitude preconditioning at 0.5\% strain per second
        \or Control fresh rat 21, with 15 cycles of 10\% strain amplitude preconditioning at 0.5\% strain per second
        \or Control fresh rat 22, with 15 cycles of 10\% strain amplitude preconditioning at 0.5\% strain per second
        \or Control fresh rat 22, with 10 cycles of 10\% strain amplitude preconditioning at 0.5\% strain per second
        \or Pregnant fresh rat 23, with 0 cycles of 10\% strain amplitude preconditioning at 0.5\% strain per second
        \or Pregnant fresh rat 24, with 10 cycles of 10\% strain amplitude preconditioning at 0.5\% strain per second
        \or Control fresh rat 25, with 5 cycles of 10\% strain amplitude preconditioning at 0.5\% strain per second
        \or Control fresh rat 25, with 2 cycles of 10\% strain amplitude preconditioning at 0.5\% strain per second
        \or Control fresh rat 26, with 0 cycles of 10\% strain amplitude preconditioning at 0.5\% strain per second
        \or Control fresh rat 26, with 0 cycles of 10\% strain amplitude preconditioning at 1\% strain per second
        \or Pregnant fresh rat 27, with 10 cycles of 10\% strain amplitude preconditioning at 0.5\% strain per second
        \or Pregnant fresh rat 28, with 10 cycles of 10\% strain amplitude preconditioning at 1\% strain per second
        \or Pregnant fresh rat 29, with 5 cycles of 10\% strain amplitude preconditioning at 0.5\% strain per second
        \or Pregnant fresh rat 29, with 10 cycles of 10\% strain amplitude preconditioning at 1\% strain per second
        \or Control fresh rat 30, with 10 cycles of 10\% strain amplitude preconditioning at 1\% strain per second
        \or Control fresh rat 30, with 2 cycles of 10\% strain amplitude preconditioning at 0.5\% strain per second
        \or Control fresh rat 31, with 10 cycles of 10\% strain amplitude preconditioning at 1\% strain per second
    \fi%
}

\newacronym{pfm}{PFM}{pelvic floor muscle}
\newacronym{pfd}{PFD}{pelvic floor disorder}
\newacronym{lac}{LAC}{levator ani complex}
\newacronym{lam}{LAM}{levator ani muscle}
\newacronym{spm}{SPM}{superficial perineal muscle}
\newacronym{spc}{SPC}{superficial perineal complex}
\newacronym{pbs}{PBS}{phosphate-buffered saline}
\newacronym{pop}{POP}{pelvic organ prolapse}
\newacronym{mri}{MRI}{magnetic resonance imaging}
\newacronym{ecm}{ECM}{extracellular matrix}
\newacronym{3ddic}{3D-DIC}{three-dimensional digital image correlation}
\newacronym{iacuc}{IACUC}{institutional animal care and use committee}

\begin{document}

\doublespacing

\maketitle

\include{Ch99_SupplementaryInfo/Abstract}


\section{Introduction}

Injury to the pelvic floor musculature represents a widespread clinical challenge that costvs the United States approximately \$50 billion USD annually \citep[cf.][]{milsom_socioeconomic_nodate,zivin_implications_2024, oneill_does_2014}. Vaginal childbirth is a significant risk factor for pelvic floor muscle injury, and approximately 85\% of women experience perineal trauma during the second stage of labor \citep{pergialiotis_risk_2014, urasaki_measurement_2023}. 
The structural failure also occurs in the long term, with roughly half of women experiencing \glspl{pfd} post-pregnancy \citep{baruch_prevalence_2023,gonzalez-timoneda_prevalence_2025}. 
Work by \citet{rieger_mechanisms_2022} shows that the pelvic floor muscles often lengthen during pregnancy, with other studies showing postpartum recovery \citep[]{suarez_pregnancy-induced_2024, boreham_appearance_2005, hsu_quantification_2006}. Studies show that in women with \gls{pop}, the \glspl{pfm} have significant shape changes and are significantly thinner when viewed with \gls{mri} \citep[]{okuda_detection_2025,hoyte_levator_2004,hoyte_biomechanics_2016}, suggesting that perhaps vaginal birth is a risk factor for prolapse because complete postpartum recovery is not always achieved.

While injuries to the pelvic floor are well documented, the underlying mechanical adaptations that occur during pregnancy are considerably more difficult to characterize \citep{alperin_pregnancy-induced_2015,rieger_mechanisms_2022,routzong_pelvic_2020}. 
Changes during pregnancy and birth allow the \gls{lam} to endure extreme strains exceeding 200\% \citep{lien_levator_2004,ashton-miller_functional_2007,delancey_pelvic_2024}, as the transverse diameter of the levator hiatus expanding from roughly 3 cm in late pregnancy \citep[]{siafarikas_levator_2014, sun_pelvic_2025}, to roughly 9-10 cm during delivery to accommodate the fetal head \citep[]{chitty_charts_1994,loughna_fetal_2009}. 
While these adaptations are necessary for successful delivery, they complicate efforts to characterize the underlying mechanical response of the pelvic floor muscles. 
Together, these challenges motivate experimental characterization of passive pelvic floor muscle mechanics that can distinguish pregnancy-related adaptation from variability introduced by the testing protocol itself.  



Experimental tests on nonpregnant tissue have provided insight into some of the mechanisms of birth injury.  
For example, in a recent study, \citet{routzong_passive_2026} measured differences in the mechanical behavior between the various muscles of the pelvic floor using female, nonpregnant human cadavers.  Their results show that the \glspl{lam} is significantly more compliant than the \glspl{spm}, for the human tissue they tested. This distinction aligns with the relative function of each tissue, where the deep \gls{lam} acs as a supportive, highly compliant sling capable of large deformation, whereas the \glspl{spm} serve a stiffer mechanical role to anchor and stabilize the pelvic outlets \citep[cf., e.g.,][]{nagle_passive_2014, janda_biomechanics_2006, vila_pouca_investigating_2020, havelkova_persistent_2020,pena_experimental_2010, collins_comparative_2026}.
The \glspl{spm} aid in anchoring the anal and vaginal openings, and 
include the perineum, which is one of the most common parts to tear during vaginal birth 
\citep[cf.][]{massalha_prevalence_2025, man_childbirth-related_2024, dendini_retrospective_nodate}.  
In contrast, the \glspl{lam}, fan out across the pelvic cavity and provide deep structural pelvic support,  
frequently experience damage, with nearly a third of women experiencing deep muscle trauma during vaginal birth \citep[]{schwertner-tiepelmann_obstetric_2012}.  Unlike \glspl{spm}, injuries to the \glspl{lam} are hidden deep within the pelvis, making them exceptionally difficult to detect and treat \citep[]{de_alba_alvarez_female_2025,estendahl_womens_2025}. The difference in stiffness of these \glspl{pfm}, as well as their locations in the body, likely explains the differing injury patterns observed between the two tissues.  
While biomechanical simulations suggest that the deeper anatomical location of the \glspl{lam} 
subject them to a different distribution of strain than the \glspl{spm} during crowning \citep[cf., e.g.,][]{routzong_pelvic_2020, jingSubjectspecificAnisotropicViscohyperelastic2012, 
vilapoucaManagementMaternalPushing2022}, its high material compliance can be pushed past its physiological limit during delivery, explaining why deep tissue failure occurs so frequently. These anatomical and mechanical complexities motivate careful experimental characterization of pelvic floor muscle behavior, particularly because measured passive properties can depend strongly on the testing conditions used to obtain them.

Few tests have been performed on pregnant \glspl{pfm} in any species.
\citet{jing_experimental_2010} performed experimental biaxial and uniaxial failure tests on vaginal tissues from pregnant rats and \gls{pfm} from squirrel monkeys and compared results to those from non-pregnant controls. 
These tests revealed that pregnancy decreases tissue stiffness in rat vaginal tissue, but the results were less clear for squirrel monkey \gls{pfm}.  Additionally, pregnancy appears to make the tissues significantly more viscous (fluid-like), allowing for better stress relaxation during birth \citep{jing_experimental_2010}.
Others have used imaging techniques to quantify non-mechanical, geometric changes to the pelvic floor \citep[cf., e.g.,][]{maranComparativeAnatomy3D2018, martinSacrumCoccyxShape2024}.

Reliable characterization of passive pelvic floor muscle mechanics is complicated by substantial experimental variability arising from tissue handling, storage conditions, loading protocols, and the inherently nonlinear, time-dependent behavior of muscle. These challenges are not unique to pelvic floor muscle and have motivated broader efforts within the biomechanics community to standardize experimental methods and reporting practices to improve reproducibility and corss-study comparison \citep[]{famaey_community_2026}.
The need for predictive modeling of the pelvic floor muscle behavior is particularly important for simulations and understanding the mechanics of \gls{pfm} injury during birth \citep[]{delancey_pelvic_2024}. 
In the studies that use cadaver tissue, the specimens often arrive frozen \citep[cf., e.g.,][]{routzong_passive_2026}. In tendon and general muscle mechanics, it is well known that freeze-thaw cycles change the mechanical properties of the tissue, often decreasing stiffness \citep[cf., e.g.,][]{fischer_influence_2020, jung_effects_2011, mcgarvey_establishing_2025}. 
Moreover, muscle in general is a complex, anisotropic, rate-dependent material, sensitive to preconditioning and loading rate \citep[]{hashemi_experimental_2020}. 
In the case of pelvic floor muscle testing, these confounding factors make it difficult to determine a baseline mechanical response of the material, even in uniaxial loading, before comparing pregnant versus nonpregnant samples.
To address this, \citet{lake_guidelines_2023} discuss standardized protocols for the mechanical testing of soft tissue.
To address these sources of variability, \citet{lake_guidelines_2023} proposed standardized approaches for mechanical testing of soft tissues, including controlled loading rates, cyclic preconditioning to establish a repeatable mechanical state, and preconditioning amplitudes selected to promote fiber alignment while avoiding tissue damage. These recommendations provide a useful framework for reducing protocol-dependent variability, although the appropriate values may depend on tissue type, scale, and mechanical behavior. While the guidelines by \citet{lake_guidelines_2023} provide a framework for achieving a fairly replicable equilibrium state of tendon prior to mechanical testing, the tendon is significantly stiffer and more ordered than \gls{pfm}. In the passive muscle realm, where tissues are often less stiff, higher compliance, and tend to `blob' up when removed from the body, there remains a need to develop procedures that lead to a fairly stable equilibrium state of the tissue before testing.  

Variation in testing protocols, noisy data, and individual diversity in biomechanics lead to notoriously difficult data analysis and model parameter identification processes. As Professor Ray Ogden recalled, analysis becomes ``unwieldy as soon as the linearity... is lost"
\citep[]{ogden_large_1972, destrade_ogden_nodate}.
\citet{destrade_methodical_2017} discuss the fitting of a model to mechanical data for rubber, and offers a way to discard artifacts of the test, such as startup actuator spikes, and ways to fit models to portions of the data within strain ranges of interest. 
High-strain or failure testing of tissues often shows micro-tearing before bulk fracture, creating a highly nonlinear bulk response \citep[cf., e.g.]{deo_multi-phase_2026, chittajallu_review_2022, lemaitre_ed_handbook_2001}. 
Therefore, data from these passive failure tests on tissues is often noisy and may include experimental artifacts, such as mini fractures and plateaus as new parts of the tissue become engaged.  These artifacts need to be  identified and accounted for in some way in the data analysis before constitutive modeling. In soft tissue mechanics, constitutive model identification/calibration is often treated as a curve-fitting exercise across population averages. However, protocol-dependent variability and experimental artifacts may lead to unrealistic constitutive models if a stable equilibrium state is not achieved or experimental conditions are not properly considered. 

To the best of the authors' knowledge, this study is one of the first direct comparison between pregnant and non-pregnant tissues, specifically in rat \glspl{pfm}. Accordingly, this study pursued two linked aims. First, we quantified how loading rate, preconditioning, storage condition, and analysis choices influence the measured passive mechanical response of rat pelvic floor muscle in order to establish a repeatable testing framework. Second, using this framework, we characterized pregnancy-associated differences in passive pelvic floor muscle mechanics. For the first aim, we systematically evaluate a range of parameters, encompassing loading rates (0.5~\%/s vs. 1~\%/s), preconditioning strain amplitudes (0\%, 2\%, 5\%, 10\%, and 15\% strain for ten cycles), and post-mortem storage states (fresh vs. frozen), to quantify their explicit effects on uniaxial high-strain behavior and failure metrics in a rat model. By tracking shifts in peak stress, failure strain, strain energy density, and tangent moduli, we isolate how altering experimental conditions skews the apparent material response. 
In particular, the threshold at which preconditioning drives the tissue into a stable, repeatable, fiber-aligned equilibrium state \citep[cf., e.g.,][]{cheng_effects_2009, wu_effects_2003, tonge_minimal_2013}, is investigated as a potential source of discrepancies in the literature. To manage the inherent numerical noise of micro-structural tearing before bulk fracture, this work establishes an objective, robust criterion for isolating the linear region of the stress-strain curve, ensuring parameter identification remains invariant to localized mini fractures and experimental artifacts. 
For the second aim, we applied the resulting experimental framework to nonpregnant and late-gestational rat \glspl{lam} to minimize aleatoric testing noise.
This allows us to precisely isolate how macro-scale gestational structural remodeling transforms the underlying constitutive response of the female pelvic floor. 

\section{Methods}

Late-gestational pregnant and non-pregnant (nulliparous) female Sprague-Dawley rats were obtained from Hilltop Lab Animals Incorporated (Scottdale, PA, USA), with the rats, on average, at 15 to 22 days of gestational age, and usage was in accordance with ARRIVE guidelines \citep[]{percie_du_sert_arrive_2020}. Rats were euthanized at the vendor facility prior to overnight shipment, either as fresh specimens or frozen. Experiments were conducted at the University of California, Riverside in the bioMechanics Experimental and Computational Health (bMECH) laboratory. Because all tissue samples were sourced post-mortem from the vendor, \gls{iacuc} approval was not needed for this study. All rats were kept at $4^{\circ}$C prior to testing, and experiments were conducted within 72 hours of euthanasia to maintain tissue viability pre-rigor mortis, as discussed by \citet{van_ee_effect_1998} \citep[cf., e.g.][]{van_loocke_validated_2006, pillet_effects_2022}. In some cases, the rats were shipped frozen; in that case, they were thawed overnight at $4^{\circ}\text{C}$.  A total of $n = 19$  fresh and $n = 2$  previously frozen non-pregnant (control) rats were utilized, and an equal number of pregnant rats were used. To preserve the native structure and prevent freeze-thaw artifacts or room-drying tissue collapse, intact whole rat carcasses were stored at 4°C immediately post-mortem \citep[]{tuttle_post-mortem_2014,safa_exposure_2017}. Extracted samples were kept hydrated with frequent sprays of 1X PBS (Corning, Manassas, VA, USA) prior to testing \citep[]{nicolle_dehydration_2010}.  

\subsection{Collection of Pelvic Floor Muscles}

\begin{figure}[htbp]
    \centering
    \includegraphics[width=1\textwidth]{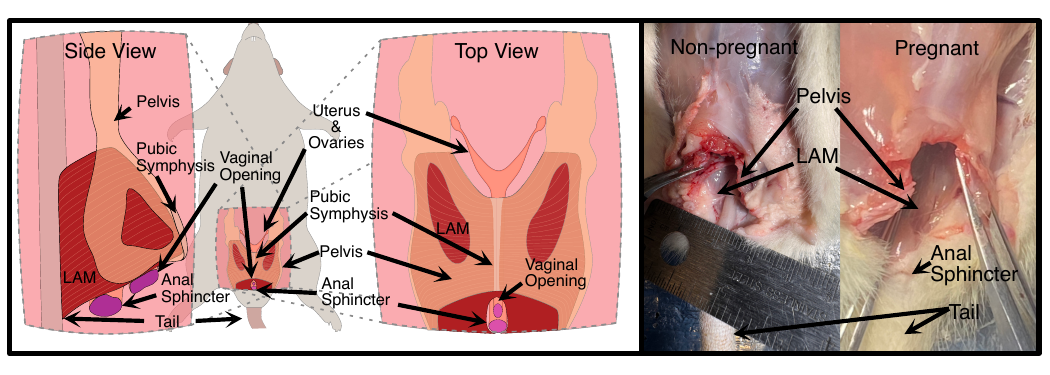}
    \caption{Diagram of the female rat pelvic floor muscles. Diagram on the left shows zoomed in side and top views of the major structures in the rat pelvic floor and the diagrams on the left show corresponding surgical images of pregnant and control rats.}
    \label{fig:FigM01_RatAnatomy}
\end{figure}

Due to the small physical dimensions of the rat pelvic floor components, the \glspl{lam}, consisting of the pubocaudalis and iliocaudalis muscles, were harvested as a single unit, and no distinction was made between these two muscles. Extraction of the pelvic floor muscles involved exposure of the pelvic cavity and manual removal of the left and right sides of the muscle. \Cref{fig:FigM01_RatAnatomy} shows a diagram of female rat anatomy along with a dissection image of the cavity created to remove the \gls{lam} before testing. Briefly, rats were splayed out on a cutting board with pins, and the skin was trimmed on the abdomen to expose the pelvic region. The anterior pelvic bone was located, and the pubic symphysis was cut through with surgical scissors. The pubic bone was gently rotated to expose the pelvic cavity, and the muscle was removed from the lining of the interior pelvic bone, and finally, as needed, any attached anal and vaginal tissues that remained attached to the \gls{lam} were severed. The left and right segments of the \glspl{lam} were removed separately, and thus, two samples per rat were obtained. Preliminary tests did not show significant differences in mechanical properties between the left and right parts of the \glspl{lam}. Immediately upon removal, any excess soft tissues were trimmed, and the resulting specimen was irregularly shaped, but on the order of 2~cm $\times$ 1~cm. Samples were mounted to a mechanical loader (BioTester 3000 with 2.5~N load cell, CellScale Biomaterials Testing, Waterloo, ON, CAN), in which they were loaded in uniaxial tension using standard metal screw grips, generally aligning with the fiber orientation in the tissue \citep[cf.][]{margulies_appearance_2006,routzong_sexual_nodate,bremer_innervation_2003}. Images were taken of the top-, front-, and back-views of the specimen with a ruler in the background, to measure thickness and width (using ImageJ, NIH, Bethesda, USA). Five measurements were taken and averaged, and measurements from both the front and back views of the specimen were used to determine thickness in order to account for through-thickness irregularities, a common occurrence in soft tissue specimens that must be accounted for \citep[]{ge_crosssectional_2020}. The calculated area and uncertainty in the area for each sample is presented in \Cref{tab:TableS01_ExperimentalConditionSummary}. 

\subsection{Mechanical Testing}
Samples were made taut by loading to approximately 2\% of the overall failure stress (as recommended by \citet{lake_guidelines_2023}), determined by the average of preliminary experiments. However, given that 2\% of the expected failure stress was generally within the precision of the load cell, for practical purposes, the specimen was considered taut when the force read by the load cell was consistently positive.  
Once the sample was made taut, images were captured to measure the specimen cross-sectional area.  
Then the sample was lowered into a 1X \gls{pbs} bath at $37^{\circ}$C, average Sprague-
 Dawley internal temperature \citep[]{suarez_mechanical_2026,mcguire_tear_2021,oliva-hernandez_single_2018,noauthor_astm_nodate}.  For a few tests, the sample remained in ambient conditions for comparison with human tissue tests by \citet{nagle_passive_2014, routzong_passive_2026}, which were also in ambient conditions.   
 The sample length was used to calculate strains and strain rates for input into the CellScale software.  This step ensured that consistent strain rates, rather than displacement rates, were applied to each specimen during loading, as discussed in work by \citet{huang_experimental_2019,nafar_dastgerdi_comprehensive_2021}.  Samples were loaded in uniaxial tension at either 0.5\% engineering strain per second or 1\% engineering strain per second for the entirety of the respective tests, as discussed in \citet{lake_guidelines_2023}. To determine a suitable loading procedure for the high-strain properties of the \glspl{lam}, following protocols adapted from \citet{nelson_human_2026} and general guidelines by \citet{lake_guidelines_2023}, a 10-cycle triangular-wave form preconditioning engineering strain of either 0\%, 2\%, 5\%, 10\%, or 15\% was applied to the sample, followed by a 15-minute rest period, before loading to failure to erase the strain history and allow for viscoelastic relaxation \citep{tonge_minimal_2013}. In \citet{carewRolePreconditioningRecovery2000}, resting the sample yields more repeatable material responses than non-rested samples; in \cite{moura_experimental_2025}, 15 minutes was sufficient to fully relax human cadaver pelvic floor muscles, specifically the perineal body. 
Axial force and displacement were recorded for each sample and used to calculate true stress and strain data, which provide more accurate representations of the local material behavior than displacement and force \citep[cf., e.g,][]{huang_experimental_2019,nafar_dastgerdi_comprehensive_2021}. 
Tests were performed on either pregnant or nonpregnant (control) specimens that arrived fresh or frozen, so that both pregnancy and tissue conditions could be compared in addition to strain-rate and preconditioning strain amplitude. 

\subsection{Data Analysis and Statistics}
\subsubsection*{Peak Stress-Strain Metrics}
All data analysis and statistical comparisons were performed in MATLAB (Mathworks, Natick, MA, USA), using the Statistics and Machine Learning Toolbox. The averaged sample cross-sectional area was used to calculate the true stress from the force response, and axial strain was calculated from the measured displacement and initial grip-to-grip length, with force and displacement zeroed post-testing. Peak stress and strain at sample failure were calculated for each sample, as shown in \Cref{fig:FigS_PeakValuesDemo}, representing the material's ultimate strength and elongation capacity. Pre-peak strain energy density was calculated as the area under the stress-strain curve up to the peak, as a measure of elastic resistance to failure (see \Cref{fig:FigS_PeakValuesDemo}). 

\subsubsection*{Toe and Linear Region Slope}
Tensile stiffness was estimated from the dominant linear region of each pre-peak stress--strain curve using an automated linear-region detection scheme, consistent with prior passive-muscle studies that quantify regional stiffness from locally linear portions of nonlinear stress--strain responses \citep[cf.,][]{brown_passive_2012,lieber_biochemical_2021,spyrou_muscle_2011}. The dominant linear region was defined as the portion of the pre-peak stress--strain curve where the local slope was most consistent, highest, and the stress--strain response was approximately linear. Details on the automated algorithm are presented in \Cref{s:data_in_boxplots} and the algorithm itself is publicly available on the \href{https://www.mathworks.com/matlabcentral/fileexchange/184316-detectlinearregion/?s_tid=mlc_lp_leaf}{MATLAB Central File Exchange}. 

To further characterize the low--strain mechanical response, transition metrics were estimated from the early portion of each stress--strain curve. For each sample, the stress-strain response was smoothed using a Savitzky--Golay filter, to reduce high-frequency noise while preserving the overall shape of the curve. Before fitting, each smoothed curve was baseline-corrected so that the first point of the analyzed curve corresponded to zero stress and zero strain. The portion of the curve below 0.1 strain was excluded from the primary transition--point search to reduce sensitivity to startup noise, low--amplitude loading artifacts, and specimen slack. 
This cutoff was supported by a startup--noise analysis in which noise was estimated as the standard deviation of the raw stress residuals relative to the smoothed stress curve over the 0--0.10 strain interval. Across samples, the median residual noise in this interval exceeded the median stress rise over the same interval, indicating that the earliest portion of the curve was dominated by measurement noise and settling behavior rather than consistent specimen loading. 
Toe-region stiffness, $E_{\mathrm{toe}}$, was estimated using a fixed-window linear-fit sweep. Candidate transition points were evaluated between 0.10 and 0.40 strain by fitting the smoothed stress--strain response from the zeroed starting point of the curve to each candidate endpoint. The candidate endpoint with the highest $R^2$ value was selected as the transition point, and the slope of this fit was reported as $E_{\mathrm{toe}}$. Transition stress and strain were defined as the smoothed stress and strain at the selected endpoint. 
If the maximum $R^2$ occurred at the lower bound of the primary search window, the search was repeated once with an expanded lower bound of 0.05 strain. If the maximum $R^2$ occurred at the upper boundary, the search was repeated once using an expanded upper bound of 0.80 strain. In either case, the highest $R^2$ endpoint from the expanded search was accepted. The selected toe--region fits and the corresponding slopes are shown with the raw stress--strain curves in \Cref{fig:FigS_SavitskyGolayPanel}. 

\subsubsection*{Hysteresis and Cycle-10 Preconditioning Loop Area}
Preconditioning cycle area was calculated for the samples that underwent cyclic preconditioning as a measure of mechanical energy loss. 
It is important to note that these cycles were before resting the material, and therefore, residual strains may be present. 
For each sample, the 10th preconditioning cycle was smoothed using a Savitzky--Golay filter and split into loading and unloading branches at the maximum strain. Hysteresis was then calculated as the area between the loading and unloading branches of the cycle \citep[]{lake_guidelines_2023,sommer_quantification_2015,quiros_comparative_2024}. Because stress was reported in kPa and strain was dimensionless, the resulting area has units of kPa, equivalent to $kJ/_{m^3}$. Samples without preconditioning were not assigned a cycle-10 loop area. 

\subsubsection*{Experimental Statistics}
With peak stress, peak strain, pre-peak strain energy density, high-strain linear stiffness, transition metrics, and cycle-10 preconditioning loop area calculated for each applicable test, group-wise comparisons were performed to evaluate the effects of pregnancy condition, strain rate, preconditioning level, hydration state, and storage condition on each mechanical measurement. Fresh samples were compared across pregnancy condition, preconditioning level, and loading rate. Fresh and frozen samples were compared at 1\% strain per second across matched preconditioning levels. Initial linear interaction models were used to evaluate global main effects and multi-factor interactions. To preserve degrees of freedom given the inherently small sample sizes, non-significant interaction terms were systematically removed by backwards elimination to yield parsimonious omnibus models \citep[Sec. 5.4.7.3]{nistsematech_nistsematech_2012}. To mitigate Type I error accumulation, the linear models were used as the primary inferential framework for evaluating global main effects and interaction s before considering follow-up pairwise comparisons \citep{schoot_small_2020,nistsematech_nistsematech_2012}. When a global main effect achieved statistical significance ($p<0.05$), followup two-tailed Welch's $t$-tests were used to describe the corresponding pairwise group-level differences \citep[cf., Sec. 7.4.7]{nistsematech_nistsematech_2012}. Because follow-up pairwise comparisons were limited to a defined set of scientifically motivated comparisons and interpreted in the context of corresponding statistically supported model terms, no additional blanket correction was applied across all pairwise tests. Multiplicity was instead considered with respect to the inferential purpose and comparison family, including whether multiple tests contributed jointly to a single confirmatory conclusion \citep[]{bender_adjusting_2001,hoffmann_when_2026}. This approach also avoids treating all reported comparisons as a single universal family, an overly conservative practice that has been criticized for unnecessarily increasing Type II error \citep[]{rothman_no_1990}. Experimental sub-tiers with sample sizes of $n=2$ were evaluated only as exploratory trends rather than to infer population-level effects. Corresponding significance bars were added to box plots for pairwise comparisons supported by the corresponding omnibus model term ($p<0.05$). To highlight potential directions for future study, borderline global trends ($0.05 \leq p \leq 0.10$) were reported in the text and treated as strictly exploratory, but were not used to justify pairwise post-hoc testing.

\subsubsection*{Literature Comparison Statistics}
To contextualize the experimental rat models within existing human clinical data, literature values for human cadaveric \glspl{lam} were digitized using PlotDigitizer (AnvSoft Inc.) from \citet{nagle_passive_2014,routzong_passive_2026} and compared against the current dataset. For this external comparison, experimental rat data evaluated at 0.5\%~s$^{-1}$ strain rate were pooled across preconditioning levels into aggregate control ($n=14$) and pregnant ($n=15$) cohorts. This pooling was used to provide a higher-level comparison of response magnitudes across pregnancy condition and species/literature source, rather than to test preconditioning-specific effects. A third unsubmerged `dry' family was added to the analysis of rat samples tested at 0.5$\%/s$ with a 2\% preconditioning strain amplitude ($n=4$) to compare unsubmerged protocols to human tissue and test the effects of hydration.
Note that henceforth, if the sample is not labeled as `dry' or noted as unsubmerged, the test was conducted in the \gls{pbs} bath.  
For each metric, group-level descriptive statistics and variance diagnostics were first evaluated. Data normality was assessed using the Lilliefors test. Given the heterogeneous data set (pooled control/pregnant rat and human digitized data) and distributional/variance characteristics of the five groups, global differences were evaluated using Kruskal-Wallis tests. When a significant global effect was identified, post-hoc pairwise comparisons were performed using Dunn--Sidak-adjusted rank-based multiple comparisons in MATLAB. These analyses were interpreted as contextual comparisons of response magnitude rather than controlled cross-species equivalence tests. 
\section{Results}
\subsection{Comparison of Preconditioning Levels and Strain Rates}

\subsubsection{Linear Model Analysis}

\begin{table}[htb]
\caption{Global main effects for high-strain tissue mechanics. Omnibus additive linear models evaluated the influence of pregnancy condition, preconditioning level, and loading rate on fresh samples. Significant effects (`*' denote $p < 0.05$) served as strict gates for subsequent pairwise comparisons. Notably, pregnancy condition emerged as the sole significant predictor, distinctly governing peak true strain.  There were several borderline significant effects, such as preconditioning and rate interacting as a significant predictor of both peak stress and stiffness.}
\label{tab:TableR01_PreconHighStrain}
\resizebox{\textwidth}{!}{%
\begin{tabular}{cccccc}
\hline
Metric                & Formula                                    & Condition & Precon & Rate & Rate:Precon \\ \hline
Peak Stress           & TrueStress $\sim$Condition + Rate + Precon & 0.97      & 0.50   & 0.90 & -           \\
Peak Strain           & Strain $\sim$Condition + Rate + Precon     & 0.02*      & 0.54   & 0.09 & -           \\
Stiffness             & Stiffness $\sim$Condition + Rate + Precon  & 0.57      & 0.09   & 0.49 & -           \\
Strain Energy Density & Wmax $\sim$Condition + Rate + Precon       & 0.69      & 0.99   & 0.68 & -           \\
Peak Stress           & TrueStress $\sim$Condition + Rate * Precon & -         & -      & -    & 0.08        \\
Stiffness             & TrueStress $\sim$Condition + Rate * Precon & -         & -      & -    & 0.06       
\end{tabular}%
}
\end{table}

To establish the overarching statistical filter prior to pairwise comparisons, additive omnibus linear models were evaluated for all primary high-strain mechanical metrics (\Cref{tab:TableR01_PreconHighStrain}). Among these global models, pregnancy condition emerged as the sole statistically significant main effect, specifically governing peak true strain ($p=0.02$). The corresponding model coefficients quantified this physical difference, indicating that the gestational state independently increased peak tissue extensibility by an average of 0.40 strain relative to controls, even after accounting for loading rate and preconditioning levels. Furthermore, the models identified additional borderline global trends ($0.05 \leq p \leq 0.10$) that were treated as strictly exploratory. First, loading rate exhibited a trending negative effect on peak strain ($p=0.09$). Model coefficients estimated a $-0.37$ reduction in peak strain when the loading rate was increased from 0.5\%/s to 1.0\%/s. Second, preconditioning level demonstrated a trending global effect on high-strain linear stiffness ($p=0.09$). In contrast, the peak true stress and the pre-peak strain energy density were entirely insensitive to the pregnancy condition, the preconditioning level, and the loading rate within the tested ranges, not producing overarching effects supported by the model. Although the additive models did not identify significant independent effects of preconditioning or loading rate on peak stress or high-strain linear stiffness, the interaction models identified borderline rate-by-preconditioning trends for both metrics. Specifically the rate-by-preconditioning interaction showed borderline effects for peak stress ($p = 0.08$) and high-strain linear stiffness ($p = 0.06$). These trends suggest that preconditioning did not influence the tensile response uniformly across loading rates, motivating closer examination of the corresponding pairwise comparisons and box-plot patterns.

\subsubsection*{Pairwise Comparisons}
\begin{sidewaysfigure}[htbp]
    \centering
    \includegraphics[width=1.0\textheight]{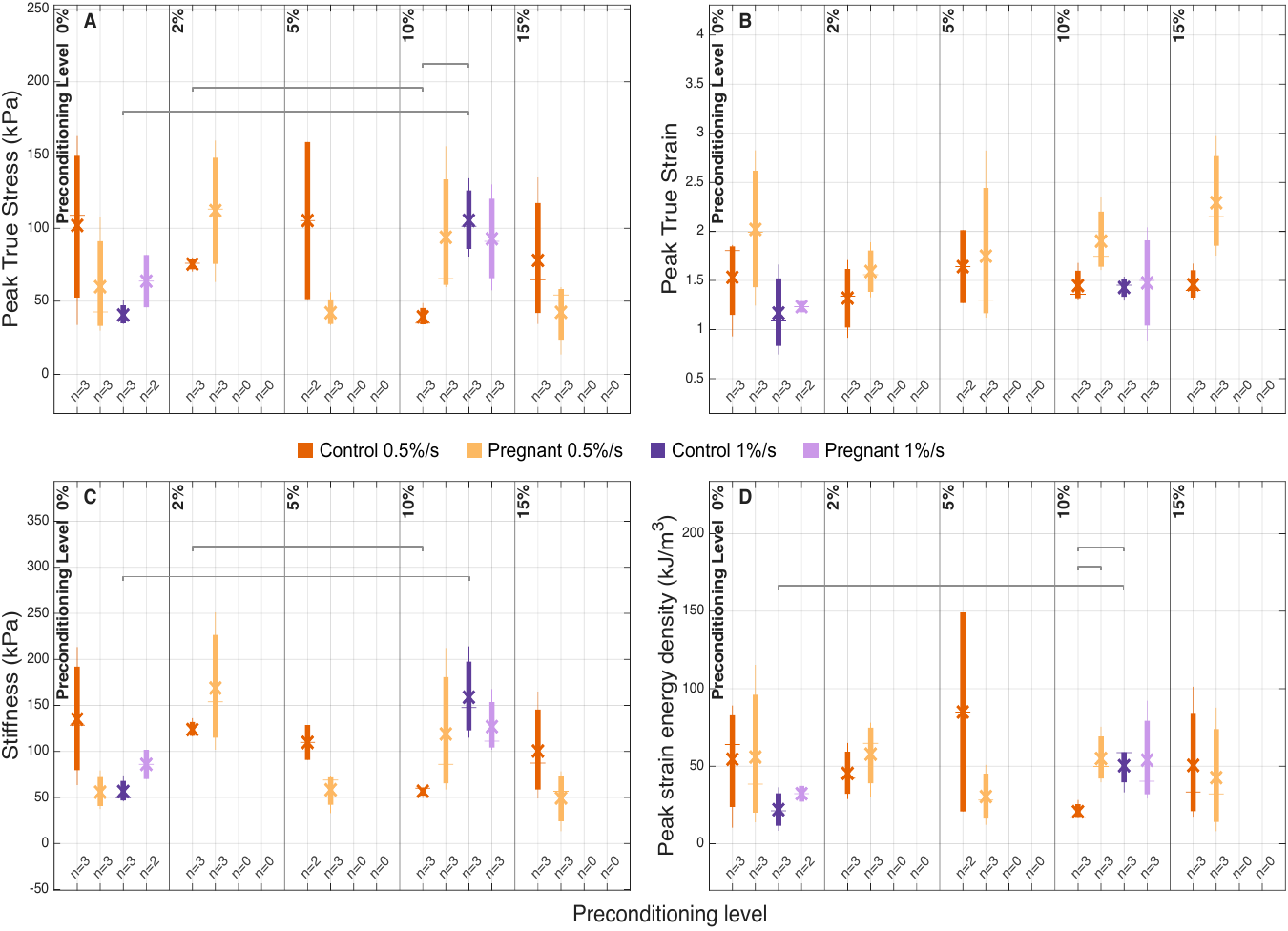} 
    \caption{Uniaxial tensile response of control and pregnant rat pelvic floor samples tested at 0.5 and 1\% strain per second across preconditioning levels. Box plots show: A, peak true stress; B, peak true strain; C, high-strain linear stiffness; and D, pre-peak strain energy density. Each panel is segmented by preconditioning strain; sample size is shown along the axis, and means are indicated by color-matched ``x'' markers. Solid and dashed brackets indicate one-factor and multi-factor comparisons, respectively. Gray brackets denote exploratory comparisons only, where Welch comparisons did not show significant trends but $p<0.10$ was nearing significance.}
    \label{fig:FigR_PreconditionComparisonBoxplotPanel}
\end{sidewaysfigure}

\Cref{fig:FigR_PreconditionComparisonBoxplotPanel} compares peak stress, peak strain, linear region stiffness, and pre-peak strain energy density across preconditioning levels for fresh samples tested at 0.5 and 1\% strain per second. Corresponding raw data curves are shown in \Cref{fig:FigS_PreconditionComparisonSSCurves}. 
Due to small sample sizes,  observations in this section are trend level, rather than definitive statistical statements. 
Box plot comparisons show exploratory significance bars, in which Welch tests showed significance ($p<0.10$) even if the linear models did not show significant trends. 

Linear model comparisons did not identify preconditioning level, loading rate, or the `pregnancy condition $\times$ rate' interaction terms as significant predictors of peak stress or stiffness trends. 
At 0.5\%/s, 
both peak stress and stiffness are either maintained or elevated when going from no preconditioning to 2\% preconditioning.  The pregnant tissues then show a decline in both stiffness and peak stress around 5\% preconditioning, followed by a rise again at 10\% preconditioning.  We think the decline may be associated with microscopic damage in the tissue \citep[cf.,][]{stauber_tendon_2020,provenzano_subfailure_2002}, while the subsequent increase may reflect greater fiber recruitment and alignment with the loading direction, as increased skeletal-muscle fiber alignment has been associated with higher passive elastic modulus \citep[cf., e.g.,][]{liang_structural_2026,wheatley_investigating_2020}.  
While the control rat tissue shows little change in stiffness or peak stress until around 10\% preconditioning strain, suggesting some difference in the onset of damage with pregnancy.  
Additionally, this decline in both peak stress and stiffness from the 2\% to the 5\% (in the pregnant case) or 10\% (in the control case) preconditioning levels for the slower loading rate suggests that the preconditioning level can affect the long-term behavior of the material. 

In contrast, although the 1\%/s groups were sparse at several preconditioning levels, the available data suggest that peak stresses may be higher at higher preconditioning levels, particularly near 10\% preconditioning strain amplitude. 
Similarly, stiffness at 1\%/s appeared to increase as the preconditioning level increased, although this trend was also limited by missing or low sample counts at some preconditioning levels. 
Again, this change in stiffness and peak stress suggests that preconditioning can affect high-strain mechanical behavior of the material.  
Moreover, the differences between the trends in stiffness and peak stress as a function of preconditioning and strain rate suggest that the preconditioning level needed to reach maximum alignment with minimal damage may depend on strain rate.

Peak strain showed significant dependence on pregnancy in the linear models; however, this is less clear in pairwise comparisons, where no significance is seen in the peak strain.  
Nonetheless, the effect is still visible when comparing trends in \Cref{fig:FigR_PreconditionComparisonBoxplotPanel}, where pregnant samples generally reached higher peak strains that matched control samples, though the magnitude of this difference varied with preconditioning level and loading rate. Peak strain also tended to be lower at the faster loading rate, consistent with a rate-dependent reduction in deformation \citep[]{liu_anisotropic_2019}. Strain energy density was more variable across groups. 
As strain energy depends on both peak stress and peak strain, it is not surprising to see that some of the significance trends observed for peak stress are the same in the peak strain energy density.  

Overall, these data suggest consistent results are achieved at low preconditioning levels for 0.5\%/s testing, whereas 1\%/s testing may require higher preconditioning levels to produce more stable responses. These protocol interpretations are trend-based, because preconditioning effects and rate-preconditioning interactions were not statistically significant in combined models. 

\subsection{Comparison of Preconditioning and Pregnancy in Toe Region}

\subsubsection*{Linear Model Analysis}

\begin{table}[htb]
\caption{Global main effects for low-strain and transition mechanics. Omnibus additive linear models evaluated the toe-region behavior of fresh samples across pregnancy condition, preconditioning level, and loading rate. Significant main effects (`*' denote $p < 0.05$) indicate that low-strain mechanics are primarily governed by rate-dependence and loading history rather than gestational state.}
\label{tab:TableR02_PreconLowStrain}
\resizebox{\textwidth}{!}{%
\begin{tabular}{cccccc}
\hline
Metric                                  & Formula                                          & Condition & Precon & Rate & Precon:Rate \\ \hline
Transition Stress                       & TransitionStress $\sim$Condition + Rate + Precon & 0.68      & 0.71   & 0.11 & -           \\
Transition Strain                       & TransitionStrain $\sim$Condition + Rate + Precon & 0.21      & 0.01*   & 0.05* & -           \\
Toe Slope                               & StiffnessToe $\sim$Condition + Rate + Precon     & 0.36      & 1.00   & 0.00* & -           \\
Cycle-10 Area (kJ/m\textasciicircum{}3) & Cycle10AreaAbs $\sim$Condition + Rate + Precon   & 0.60      & 0.01*   & 0.26 & -           \\
Transition Strain                       & Stiffness $\sim$Condition + Rate * Precon        & -         & -      & -    & 0.06 
\end{tabular}%
}
\end{table}

In contrast to the high-strain metrics, omnibus modeling of the toe-region and transition mechanics revealed that behavior at low strains was predominantly governed by rate-dependent and loading-history (preconditioning), rather than pregnancy condition, as shown in \Cref{tab:TableR02_PreconLowStrain}. Loading rate emerged as a highly significant main effect for toe-region stiffness ($p < 0.001$). Model coefficients quantified this rate-dependent response, demonstrating that increasing the strain rate from 0.5\%/s to 1.0\%/s uniformly elevated the toe-region linear slope by 10.0 kPa. Loading rate additionally influenced the transition point ($p = 0.05$), with the faster 1.0\%/s rate causing the tissue to transition into the high-strain linear region earlier (an average reduction of 0.10). Furthermore, the level of preconditioning significantly modulated both the transition strain ($p = 0.01$) and the cycle-10 preconditioning loop area ($p < 0.05$). Coefficient estimates indicated that preconditioning the tissue to higher magnitudes (\textit{e.g.,} 10\% strain) delayed the onset of the linear region, shifting the transition point outward by 0.13 relative to un-preconditioned baselines. As expected, greater preconditioning targets also increased hysteresis, 
since more strain will lead to more area under the stress-strain curve.  
Notably, pregnancy condition had no model-supported effect on any low-strain metric, and transition true stress remained entirely invariant across all tested factors, validating its exclusion from subsequent pairwise post-hoc testing. Additionally, the interaction of preconditioning and rate before backward model elimination showed a near statistically significant trend, suggesting that the preconditioning effect was rate dependent with respect to the transition strains.  
Finally, as seen in the high-strain material behavior, these linear model trends, while depicting broad behavior, do not fully capture the pairwise reality. 

\subsubsection*{Pairwise Comparisons}

\begin{sidewaysfigure}[htbp]
    \centering
    \includegraphics[width=0.9\textheight]{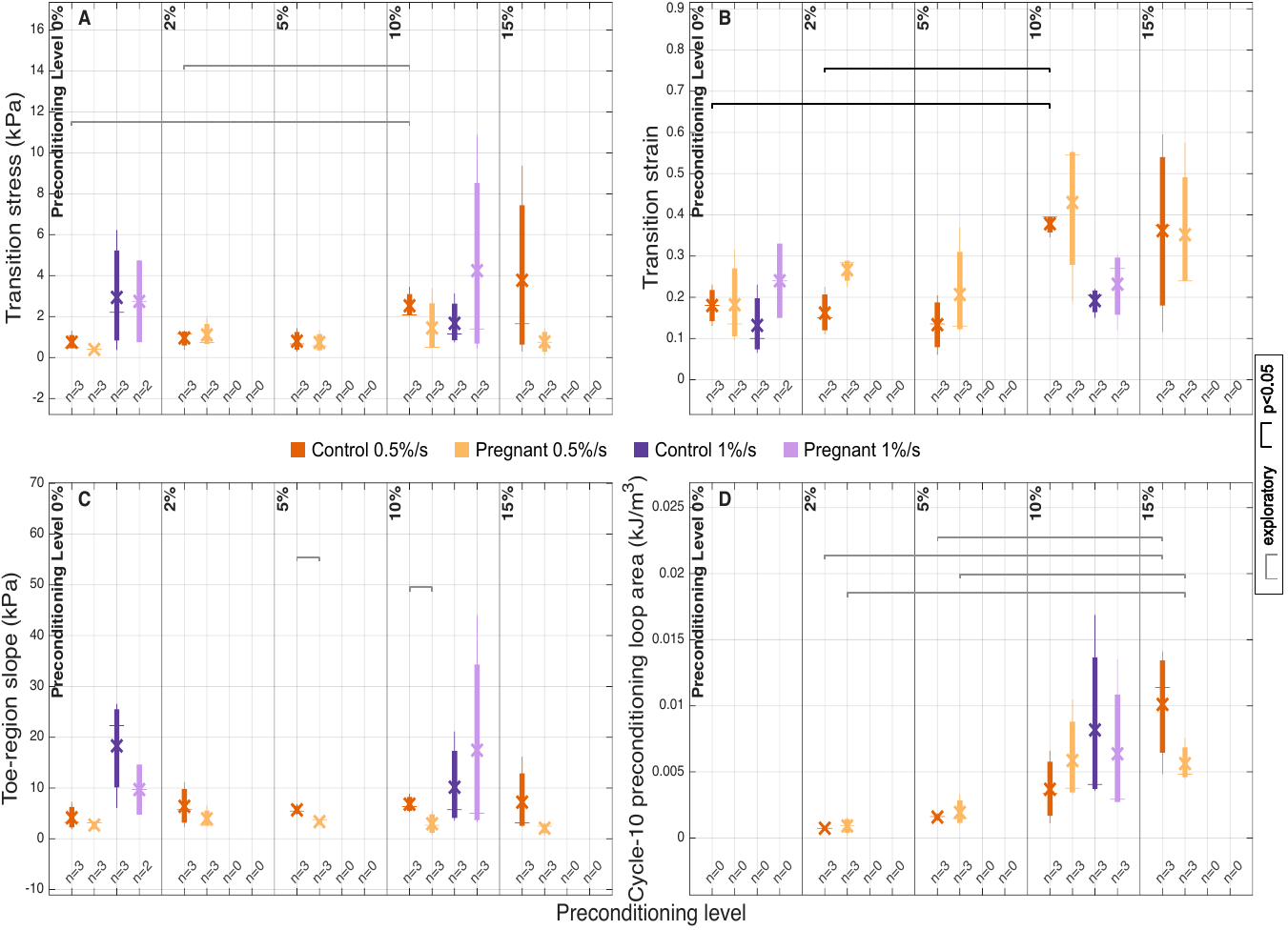} 
    \caption{Uniaxial tensile response of control and pregnant rat pelvic floor samples tested at 0.5 and 1\% strain per second across preconditioning levels. Box plots show: A, transition stress; B, transition strain; C, toe-region slope; and D, cycle-10 preconditioning loop area. Each panel is segmented by preconditioning strain; sample size is shown along the axis, and means are indicated by color-matched ``x'' markers. Black brackets denote model-supported pairwise Welch comparisons with $p<0.05$, whereas gray brackets indicate exploratory pairwise comparisons ($p < 0.1$) shown for trend-wise descriptive context.}
    \label{fig:FigS_ToeComparisonBoxplotPanel}
\end{sidewaysfigure}

\Cref{fig:FigS_ToeComparisonBoxplotPanel} shows the comparison of the transition stress and strain, toe-region slope, and preconditioning 10th-cycle area for the fresh samples at both 0.5 and 1\% strain per second at all preconditioning amplitudes. From panel A, we can see that the control samples with no preconditioning, loaded at 0.5\% strain per second, had a lower transition stress than their 10\% preconditioning-strain counterpart. In terms of the transition strains, again, there were significant differences between the control no preconditioning and the 10\% preconditioning cases, with the latter showing a longer toe region. 
This is in line with work by \citet{lee_investigating_2017} showing that micro slippage in tendon, though different than muscle, can reach local strains that break the cross-linking fibers within the material, which may make it more difficult for the fibers to become engaged. 
Similar trends are seen in the pregnant case, although they are not significant.  
Recall that the 10\% preconditioning level showed a decline in the peak stress, which may be associated with microscopic damage, so perhaps this microscopic damage may also make it more difficult to engage the fibers during loading.  

In terms of toe region slope comparisons, loading generally corresponded to higher fitted toe-region slope, particularly in the 1.0\%/s group. Several pairwise comparisons also suggested lower toe-region slopes in pregnant samples than in matched controls. However, because pregnancy condition was not identified as a significant main effect in the corresponding linear model, these pregnant-versus-control differences were interpreted as exploratory rather than as model-supported evidence of a pregnancy effect. This exploratory trend may nevertheless be consistent with pregnancy-associated remodeling of the extracellular matrix, which could allow easier fiber or matrix reorganization during the early low-strain response \citep[]{catanzarite_pelvic_2018, alperin_pregnancy-induced_2015}.

The cycle-10 area showed expected preconditioning amplitude increase, as higher strain loading leads to larger hysteresis area. 
Additionally, this increase in the cycle-10 area may also be associated with residual stresses from the previous cycles of loading. There did not seem to be apparent differences in hysteresis area in the pregnant versus control cases.

\subsection{Comparison of Storage Condition}

\subsubsection*{Linear Model Analysis}

\begin{table}[htb]
\caption{Global main effects evaluating cryo-preservation artifacts. Omnibus additive linear models compared fresh and frozen samples at a matched 1.0\%/s loading rate across preconditioning levels. Significant main effects ($p < 0.05$) driving variance in this subset are denoted by `*' and are primarily associated with preconditioning level.}
\label{tab:TableR03_StorageComparison}
\resizebox{\textwidth}{!}{%
\begin{tabular}{cccccc}
\hline
Metric                & Formula                                      & Condition & Frozen & Precon & Condition:Frozen \\ \hline
Peak Stress           & TrueStress $\sim$Condition + Frozen + Precon & 0.25      & 0.37   & 0.04*   & -                \\
Peak Strain           & Strain $\sim$Condition + Frozen + Precon     & 0.40      & 0.96   & 0.29   & -                \\
Stiffness             & Stiffness $\sim$Condition + Frozen + Precon  & 0.07      & 0.31   & 0.02*   & -                \\
Strain Energy Density & Wmax $\sim$Condition + Frozen + Precon       & 0.94      & 0.41   & 0.21   & -                \\
Peak Stress           & TrueStress $\sim$Condition * Frozen + Precon & -         & -      & -      & 0.09             \\
Stiffness             & Stiffness $\sim$Condition * Frozen + Precon  & -         & -      & -      & 0.05            
\end{tabular}%
}
\end{table}

To evaluate the potential effects of cryo-preservation on tissue mechanics, additive omnibus models compared fresh and frozen samples tested at a 1.0\%/s strain rate, as shown in \Cref{tab:TableR03_StorageComparison}. The overarching models confirmed that storage condition (fresh versus frozen) had no statistically significant effect on any primary mechanical metric, including peak stress, strain, stiffness, and pre-peak strain energy density. This null result mathematically validates that the freezing protocol did not systematically compromise or alter the tissue's tensile behavior. Instead, variance in this testing subset was primarily driven by applied preconditioning level. Omnibus models indicated that preconditioning significantly governed both peak true stress ($p = 0.04$) and high-strain linear stiffness ($p = 0.02$). Extracted model coefficients demonstrated that preconditioning the tissue to 10\% strain elevated the ultimate peak stress by an average of 50.7 kPa and stiffened the linear region by 77.8 kPa relative to un-preconditioned baselines. Additionally, pregnancy condition exhibited a borderline exploratory trend for high-strain stiffness ($p = 0.07$), with model coefficients estimating a $-33.0$ kPa softening effect in gestational state. Finally, the interaction of pregnancy condition on peak stress and linear region stiffness seemed to have near-significant p-values, suggesting that whether the rat was pregnant and arrived frozen or not may have affected the peak stress and stiffness. 
Because no effects of storage condition were supported by the global models, subsequent pairwise post-hoc tests between fresh and frozen tiers were considered purely exploratory. 

\subsubsection*{Pairwise Comparisons}

\begin{sidewaysfigure}[htbp]
    \centering
    \includegraphics[width=1.0\textheight]{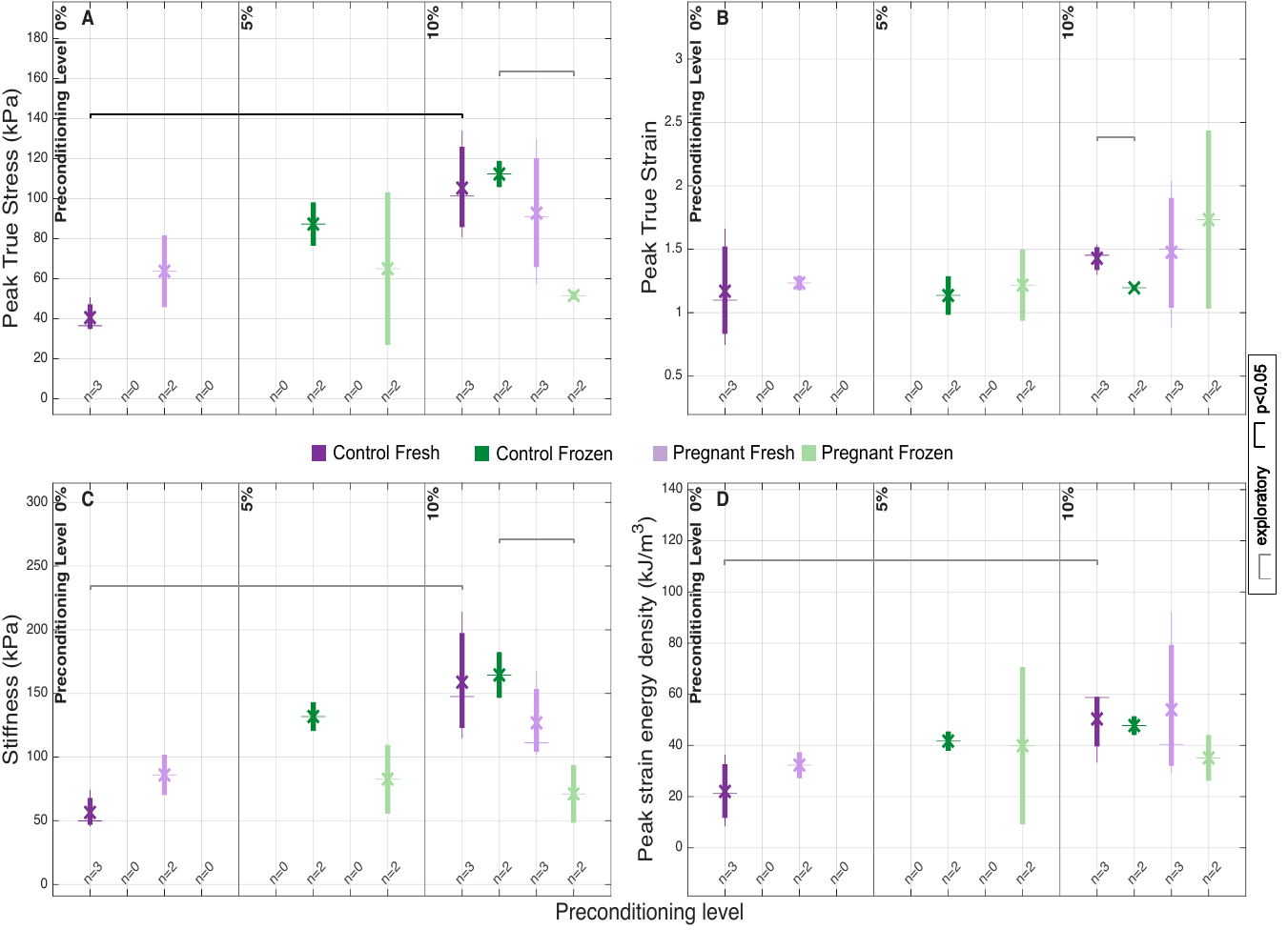} 
    \caption{Uniaxial tensile response of control and pregnant rat pelvic floor samples tested at 0.5 and 1\% strain per second across preconditioning levels. Box plots show: A, transition stress; B, transition strain; C, toe-region slope; and D, cycle-10 preconditioning loop area. Each panel is segmented by preconditioning strain; sample size is shown along the axis, and means are indicated by color-matched ``x'' markers. Black brackets denote model-supported Welch comparisons with $p<0.05$, whereas gray brackets denote exploratory comparisons only and were not interpreted as statistically significant.}
    \label{fig:FigR_FreezingComparisonBoxplotPanel}
\end{sidewaysfigure}

\Cref{fig:FigR_FreezingComparisonBoxplotPanel} compares peak stress, peak strain, dominant loading 
stiffness, and pre-peak strain energy density between fresh and frozen samples tested at 1\% strain 
per second. Corresponding raw data curves are shown in \Cref{fig:FigS_FreezingComparisonSSCurves}. 

These box plot distributions showed overlapping fresh and frozen peak stress and stiffness values across tested preconditioning levels. However, there is a statistically significant difference between the maximum stress of the fresh control without preconditioning and the fresh control at 10\%  preconditioning, and a similar nearly significant trend is seen in the stiffness. 
This follows the aggregate trend that increasing the preconditioning amplitude increases the peak stress and stiffness, particularly at higher strain rates, though, as shown in \Cref{fig:FigR_PreconditionComparisonBoxplotPanel}, there are competing mechanisms as to how this plays out. 
This is more apparent in  \Cref{fig:FigR_FreezingComparisonBoxplotPanel} as compared to \Cref{fig:FigR_PreconditionComparisonBoxplotPanel}, since there is less scarcity of data due to additional frozen samples at 5\% preconditioning strain amplitude.   
While the samples tended to get stiffer and have larger peak stress as preconditioning increased, there is an important exception with the pregnant frozen case at 10\% preconditioning strain.   
This suggests some damage in the pregnant frozen tissue that is not present in the pregnant fresh tissue, which is also suggested by the nearly significant interactions of frozen and pregnancy on both peak stress and stiffness.    

Peak strain also did not show a statistically significant storage effect in the linear models. Nevertheless, the raw curves and box plot distributions suggest 
some potential changes in deformation behavior. 
There is borderline significance between the peak strain of the frozen and fresh control samples at 10\% preconditioning, suggesting that freezing the sample may lead to less elongation. 
Strain energy density was variable across groups and did not show significant storage effects, but may show preconditioning effects. 

Overall, these data suggest that freezing may preserve the primary load-bearing strength and stiffness in control samples, but not in pregnant samples. Additionally, these data reinforce that more preconditioning may be needed for samples loaded at a faster strain rate.

\subsection{Comparison to Literature}
\begin{sidewaysfigure}[htbp!]
    \centering
    \includegraphics[width=1.0\textheight]{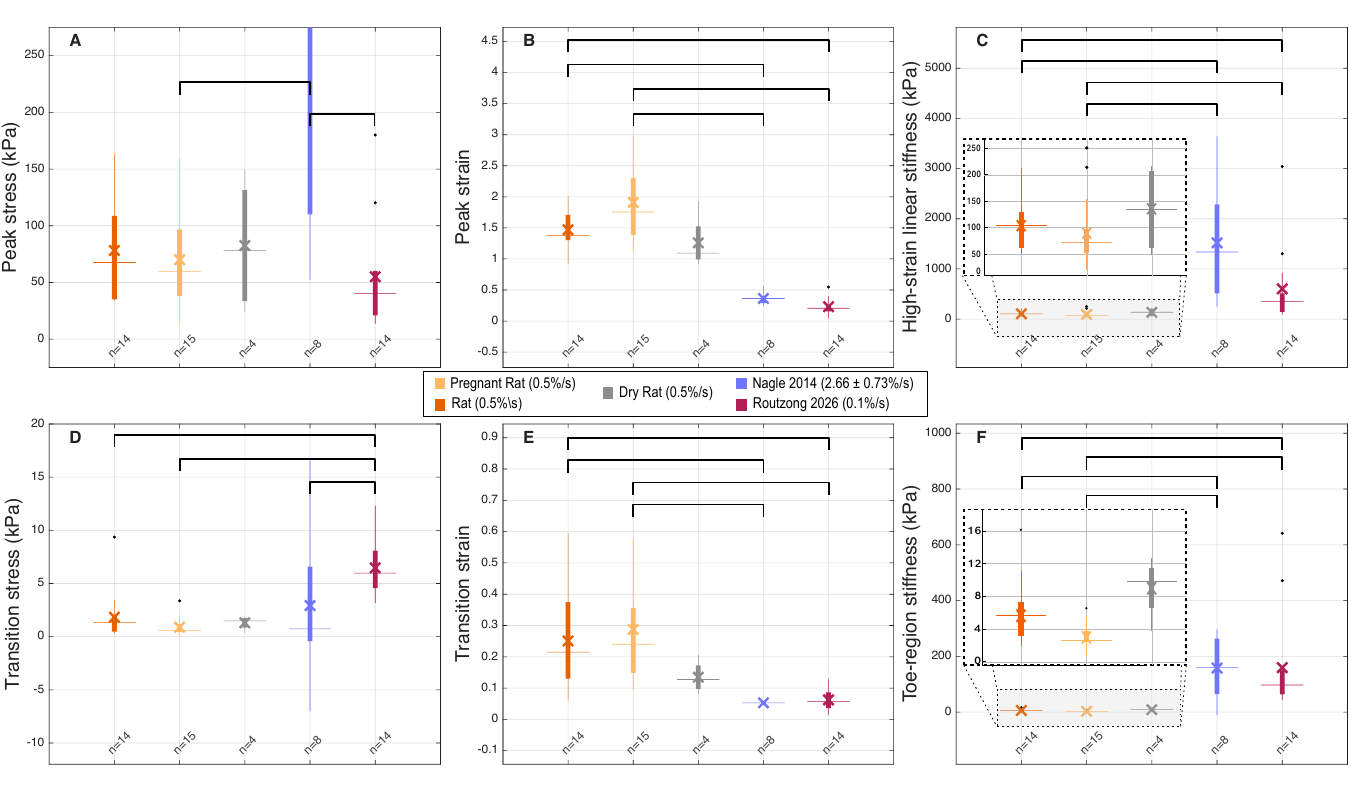} 
    \caption{Rat pelvic floor muscle data compared with digitized human cadaveric \gls{lam} data from \citet{nagle_passive_2014} and \citet{routzong_passive_2026}. To facilitate literature comparison, experimental rat data tested at 0.5\%~s$^{-1}$ were pooled across all preconditioning levels within control and pregnant groups. A third rat group of unsubmerged samples were tested at 0.5\%~s$^{-1}$ to compare to the (unsubmerged) human. Box plots show: A, peak stress; B, peak strain; C, high-strain linear stiffness; D, transition stress; E, transition strain; and F, toe-region stiffness. Sample size is indicated along the axis, and means are represented by color-matched ``x'' markers. Horizontal brackets indicate significant post-hoc pairwise comparisons following Kruskal--Wallis testing ($p<0.05$). Insets in panels C and F show zoomed-in views of the experimental rat data.}
    \label{fig:FigR_LitComparison2}
\end{sidewaysfigure}

\subsubsection*{Comparison of Protocols and Digitization of Data}
\Cref{fig:FigR_LitComparison2} compares the rat pelvic floor muscle data with digitized human cadaveric \gls{lam} data from \citet{nagle_passive_2014} and \citet{routzong_passive_2026}.
While there were important differences in the testing protocols for each case, this comparison provides literature-based context for the magnitude and range of the \gls{lam} mechanical responses and helps contextualize the rat model relative to human tissue. Unlike the experimental protocol comparisons, the literature comparison was not treated as a factorial design because species, protocol, and literature source were confounded across groups. Therefore, pooled rat control, pooled rat pregnant, unsubmerged (dry) rat (tested at $0.5\%$ and $2\%$ strain amplitude preconditioning), \citet{nagle_passive_2014}, and \citet{routzong_passive_2026} datasets were compared as five independent groups using Kruskal--Wallis tests with adjusted rank-based pairwise comparisons.
Pooling seemed reasonable given the similarity of the mechanical behavior of all preconditioning levels for pregnant/control samples, and pooling results gave a larger sample size with more statistical power.  

A detailed comparison of methodological differences between the present study and those digitized from literature is provided in \Cref{tab:TableS_MethodsComparison_Literature}. Species and tissue preparation differed substantially: the current study tested entire rat \glspl{lam}, where \citet{nagle_passive_2014} tested human levator ani segments of the iliococcygeus region, and \citet{routzong_passive_2026} tested human \glspl{lam} specimens isolated from \textit{en bloc} pelvic floor complexes. The current rat experiments also used 10 triangular-wave preconditioning cycles at prescribed amplitudes with a 15 min. rest, while \citet{nagle_passive_2014} used 10 sinusoidal cycles with 2.54 mm (approximate engineering strain amplitude of $6.4 \pm 1.7\%$ based on the reported specimen lengths) amplitude, and \citet{routzong_passive_2026} did not report preconditioning protocols. Regarding hydration, 
\citet{nagle_passive_2014} tested human tissues in ambient air ($22^\circ \text{C}$) within four hours of extraction, and \citet{routzong_passive_2026} omitted active hydration as well 
during testing to prevent interference with \gls{3ddic} strain tracking. While most of the rat tissues in the present study were tested fully submerged in a temperature--controlled bath ($37^\circ \text{C}$) $1 \times$ PBS to maintain physiological hydration, to compare rat and human tissue results a few unsubmerged samples were tested in ambient conditions so that the effect of submersion could be isolated and differences between the rat and human \gls{lam} mechanics could be compared more directly. 

Loading rates were also varied: this study applied specimen-specific, dynamically regulated strain rates (0.5 or 1\% strain per second), whereas the human literature utilized fixed crosshead displacement rates of 1 mm/s, which corresponds to $2.66 \pm 0.73$\% per second  \citep[]{nagle_passive_2014}, and 5 mm/min, which corresponds to 0.1\% per second \citep[]{routzong_passive_2026}. 
When comparing the results for rat \gls{lam} mechanics to human \gls{lam} mechanics, we only considered fresh tissue at the 0.5\% strain per second strain rate.  Only one strain rate was chosen because, as \Cref{fig:FigR_PreconditionComparisonBoxplotPanel} shows, strain rate affects the mechanical behavior of the material, and this slower rate most closely matches the strain rate by \citet{routzong_passive_2026}. 

During digitization of the reported stress-strain curves for these studies, the reported lengths for each sample and cross-sectional gauge area were used to convert force-displacement data to stress-strain data \citet{nagle_passive_2014, routzong_passive_2026}. Although both \citet{nagle_passive_2014} and \citet{routzong_passive_2026} included stress--strain data, literature curves were reconstructed from digitized force--displacement data so that stress and strain could be calculated consistently using the reported specimen geometries. This avoided mixing different or unspecified strain and stress conventions across datasets. 
Digitized data were analyzed
using the same pipelines as the raw experimental data from the current study, with a minor
modification. The noise filtering was not used on the digitized data for linear and toe analysis, due to the inherent smoothness of the digitized curves. Thus, trends should be interpreted with caution, as values for some metrics, particularly toe and linear region stiffness and transition stress/strain, may not have been identified exactly.  


\subsubsection*{High-Strain Data Comparison}
The top row of \Cref{fig:FigR_LitComparison2} compares the high-strain and failure data from this work to that from \citet{nagle_passive_2014} and \citet{routzong_passive_2026}.
The clearest differences between rat and human datasets were observed in peak strain. Rat samples, including both fresh and frozen groups, generally reach higher strains than digitized human \gls{lam} datasets, exhibiting consistently greater deformation capacity than the available human data. 
This may have been due to hydration during testing, as the PBS bath used in this work likely lubricated the deformation process. This is emphasized as the unsubmerged rats were not significantly different than either submerged rat or human tissue in all metrics, suggesting that dry samples tend toward stiffer material responses similar to the human tissue.  This suggests that hydration may be important to the peak strain response, but may not fully explain all differences observed between the rat and human tissues. The peak stresses for the rats were similar to those found by \citet{routzong_passive_2026}, although the \citet{nagle_passive_2014} dataset showed substantially greater variability and higher upper range values. This may be associated with the variability in loading rate and the generally faster loading rate of \citet{nagle_passive_2014} than that used in this work and in \citet{routzong_passive_2026}. 
With a faster loading rate, stresses do not have time to relax, and thus are expected to be higher.  
Similarly, human stiffness values from \citet{nagle_passive_2014,routzong_passive_2026} were generally higher and more variable than the rat values. With a faster strain rate used in \citet{nagle_passive_2014}, the load-bearing fibers do not have time to relax, and thus, stresses are expected to be higher, which would also translate to higher stiffness \citep[cf.][]{moura_experimental_2025}. 
However, the similar strain rates used between rat and human \citep[]{routzong_passive_2026} 
still yield higher stiffnesses for the human tissue, which may suggest species differences between groups or hydration differences, as the less lubricated human tissue experienced lower strains, and thus higher stiffnesses are required for similar stresses.

\subsubsection*{Low-Strain Data Comparison}
Panels D-F in \Cref{fig:FigR_LitComparison2} show the comparisons for the low-strain data. 
The transition stress, strain, and toe-region stiffness between the present study and the human digitized data show differences that may be attributed to species-to-species or testing variation. 
Transition stresses for rat samples were significantly lower than those in the data from \citet{routzong_passive_2026}, though not significantly different from those in \citet{nagle_passive_2014}. 
Both \citet{routzong_passive_2026,nagle_passive_2014} showed lower transition strains from the rat experimental groups \citep[]{zwambag_characterization_2019,burgio_mechanical_2023}. The \citet{routzong_passive_2026} data are significant because the loading rates were comparable to those in the present study, highlighting further low-strain differences between species, though the host of other factors outlined in \Cref{tab:TableS_MethodsComparison_Literature} add confounding factors.  
The toe stiffnesses between pregnant and control samples did not differ significantly; however, the rat toe stiffness from the submerged group was an order of magnitude less stiff than the human counterparts, while being similar in magnitude to the unsubmerged rat. These low-strain differences may suggest species-dependent behavior; however, the shorter/stiffer toe region in humans as compared to rats may also be associated with the hydration differences, as the fluid should allow for more slippage of molecules with respect to one another, demonstrated by the some of the similarities of the unsubmerged rat and human \citep[cf., e.g.,][]{carvalho_lose_2023,xu_understanding_2013,pourbavarasad_experimental_2026}, a phenomenon exhibited in engineered materials as well \citep[]{fellows_understanding_2020}.
\section{Discussion}

This study provides one of the first direct comparisons of passive uniaxial tensile behavior between pregnant and non-pregnant pelvic floor muscles using excised rat tissue. This study also establishes a repeatable methodology for quantifying linear-, and toe-region stiffness and transition stress/strain from nonlinear, regionally variable stress-strain curves. By varying preconditioning level, loading rate, pregnancy condition, and storage condition, these experiments demonstrate that pelvic floor muscle tensile behavior depends not only on biological state, but also on the testing conditions imposed before sample failure. 
While the differences between the mechanical response of pregnant and control tissue are relatively small, 
noticeable swelling of the material, which was tested in 1$\times$ PBS, may have contributed to this. 
Pregnancy is known to increase the fluid content of muscles \citep{lactationBodyCompositionChanges1990}, but 
if this fluid content affects the muscle mechanics cannot be assessed through our experimental results, as all pregnant samples were submerged. Future work should also consider hydration as an important testing condition, as it can affect the outcomes of mechanical tests \citep[cf.,][]{safa_exposure_2017}.  
The data also suggest that preconditioning requirements may depend on loading rate, with low preconditioning appearing sufficient at 0.5\% strain per second and higher preconditioning appearing more appropriate at 1\% strain per second. 
Overall, these results support the need to treat preconditioning, loading rate, pregnancy condition, storage history, and hydration state as important factors when designing passive pelvic floor muscle testing protocols.

\subsection{Protocol Recommendations}
The testing conditions in this study were designed to determine how passive tensile testing protocols influence the measured behavior of pelvic floor muscle, particularly when these data are intended for high-strain constitutive model calibration or computational simulations. 
Overall, findings indicate that loading rate, preconditioning history, and storage condition should be reported and selected deliberately when comparing pelvic floor muscle mechanics across studies.

\subsubsection{Preconditioning and Strain Rate}

In passive soft tissue testing, preconditioning is commonly used to reduce history dependence and improve repeatability \citep[]{carewRolePreconditioningRecovery2000}. However, the appropriate preconditioning level may depend on loading rate, tissue condition, and the strain range of interest. Moreover, the initial state of the tissue when removed from the rat was vastly different than its \textit{in vivo} analog. In the body, the bone and tissue-tissue attachments keep the material pre-loaded.  
With thees attachments severed, the tissue tended to `blob' up, which also explains why preconditioning is needed. In arterial mechanics, this issue is often framed as various types of pre-stressing of arteries that will release once the tissue is excised \citep[cf.,][]{tagiltsev_geometrically_2021,chuong_residual_1986}. 
\citet{tagiltsev_geometrically_2021} discuss how soft tissue testing involves mechanical loading of an object that has changed shape and configuration from its initial shape in the body. Preconditioning is used to return a portion of that \textit{in vivo} behavior to the tissue prior to testing \citep[]{carewRolePreconditioningRecovery2000}.


In constitutive formulations of anisotropic rate-dependent materials, such as in work by \citet{liu_anisotropic_2019} and \citet{chittajallu_review_2022}, the hyperelastic and viscous behaviors of the fibers and the \gls{ecm} are treated as distinct, decoupled contributions to the overall strain energy. Under this framework, fibers experience localized viscous slippage relative to the surrounding \gls{ecm} \citep[]{wohlgemuth_alignment_2023}, a motion constrained by structural cross-linking. This distinct matrix and fiber behavior can also be used to explain what might be occurring micro-structurally during preconditioning to explain the macroscopic behavior observed. When the structural supports of the body are removed, the fibers can relax and realign through viscous slippage, as shown by \citet{alperin_pregnancy-induced_2015}.  
Preconditioning may be particularly important when characterizing pregnant \glspl{pfm}, as \citet{alperin_pregnancy-induced_2015} demonstrated a decrease in enzymatic cross-linking of some birth canal structures during pregnancy, which may lead to additional relaxation for pregnant tissue.  
During preconditioning, the goal is to realign the fibers as they were in the body.  
\citet[]{tonge_minimal_2013} suggest that preconditioning is particularly important when characterizing the uniaxial behavior of tissues, since the fibers must be aligned to characterize the mechanics of the material in the fiber direction. 

\Cref{fig:FigS_PeakCyclic} shows that regardless of loading rate, peak-to-peak stress tends to equilibrate after ten cycles, allowing for a stable initial state of the tissue prior to testing.  Mechanistically, the initial 1-2 cycles of the preconditioning may be dominated by transient micro structural reorganization, including fiber alignment. The subsequent equilibrium and tapering observed in cycles 5-7 in \Cref{fig:FigS_PeakCyclic} represent the stabilization of the internal dissipative state into an equilibrium closed cyclic trajectory with the fibers as much aligned as they will get for the given preconditioning level \citep[]{carewRolePreconditioningRecovery2000}. The 15-minute rest post-preconditioning then allows for this aligned equilibrium state to relax and un-crimp in an aligned state, allowing for more repeatable testing, regardless of the species, loading rate, or material storage. This is discussed in work by \citet{hromada_strain-induced_2024}, where fibrous tissue scaffolds, which behave similarly to muscle, tend to align visibly after cyclic loading \citep[cf., e.g.,][]{bol_mechano-geometrical_2020,tomasch_cyclic_2023}.

The level of preconditioning contributes to the high-strain behavior after the rest period. With higher preconditioning amplitudes, more fibers are expected to align with the direction of loading, which is expected to lead to higher stiffness (as fibers are stiffer than the matrix) and peak stress (as it is more difficult to tear fibers).  However, too much preconditioning can damage the tissue and result in less stiffness and peak stress \citep[]{tonge_minimal_2013}.  
These competing factors may explain some of the inconsistencies in the trends and why pairwise and global comparisons were not significant.  

In \Cref{fig:FigR_PreconditionComparisonBoxplotPanel}, samples tested at 0.5\% strain per second showed non-monotonic changes in peak stress and stiffness across preconditioning amplitudes, with significance values for the linear models shown in \Cref{tab:TableR01_PreconHighStrain}. These trends suggest that preconditioning did not act as a simple dose-dependent effect. Instead, the response may reflect interwoven mechanisms of fiber recruitment and damage. 
Low-amplitude preconditioning may improve fiber alignment or recruitment before failure, whereas larger preconditioning amplitudes may also introduce micro-structural disruption or damage that reduces subsequent load-bearing. This interpretation is consistent with work by \citet{lee_investigating_2017}, showing that mechanically induced damage can later affect macroscopic tensile behavior through micro-structural mechanisms. In tendon fascicles, non-recoverable interfibrillar sliding has been associated with decreased linear modulus and an elongated toe region \citep[]{schatzmann_effect_1998}, indicating that the changes in stiffness and transition-region behavior can reflect altered micro-structural load transfer. Therefore, the decrease in peak stress/stiffness in pregnant samples at 5\% preconditioning followed by the increase at 10\%, and the decrease in control samples at 10\% followed by partial recovery at 15\%, may reflect concurrent effects of fiber alignment, recruitment, and preconditioning-induced micro-structural disruption. The dominant effect likely depends on pregnancy condition and preconditioning amplitude, as well as individual variations.

In contrast, considering the data in both \Cref{fig:FigR_PreconditionComparisonBoxplotPanel} and \Cref{fig:FigR_FreezingComparisonBoxplotPanel}, at 1\% strain per second, stiffness and peak stress tended to increase as preconditioning amplitude increased up to 10\%, suggesting that larger preconditioning amplitude may be more effective at recruiting or aligning load-bearing fibers at the faster loading rate, and that micro-structural slippage/damage did not have sufficient time to occur at this loading rate, though more higher preconditioning amplitude responses need to be sampled to provide a definitive statement. Together, these results indicate that the mechanical effects of preconditioning were non-monotonic and likely depended on the balance between fiber alignment, viscoelastic relaxation, and preconditioning-induced tissue disruption \citep[]{lee_investigating_2017,provenzano_subfailure_2002}. 

The connection between loading rate and preconditioning level is likely associated with the viscous relaxation that occurs with slower tests, as discussed in \citet{rehorn_passive_2014}.  As the material relaxes, the fibers can align, even with less initial load. Linear models of transition-region metrics suggest that preconditioning affected the strain at which the low-strain response transitioned toward the higher-stiffness region, but had less consistent effects on the toe-region stiffness itself. 
Specifically, the transition strain shifted higher for higher preconditioned samples. This suggests that preconditioning may alter how readily the tissue can be straightened, aligned, or damaged before fiber recruitment, while the apparent toe-region stiffness may remain more strongly governed by the underlying tissue structure. This interpretation is in keeping with work by \citet{schatzmann_effect_1998}, which discusses the idea that preconditioning may have competing effects in tissue organization. Initial loading may help align or recruit longer intact fiber bundles, analogous to pulling long pieces of spaghetti straight, whereas larger or repeated loading may introduce disruption (like cutting the spaghetti) that makes subsequent alignment less efficient. Thus, changes in transition strain may reflect the amount of deformation required before the tissue reaches a more load-bearing configuration while toe-region stiffness may reflect the intrinsic resistance of the low-strain fiber or matrix-dominated response \citep[cf.,][]{bianco_effect_2021,connizzo_structurefunction_2013}.

Additionally, linear model results strongly suggest that the loading rate also affects the toe-region slope, a trend observable in \Cref{fig:FigS_ToeComparisonBoxplotPanel}, with increased loading rate increasing toe-region slope, though pairwise comparisons suggest competing factors that may confound significant observations. This is again likely associated with viscous relaxation and larger stresses for the same strain without time for relaxation.  

Added together, these results suggest that preconditioning may have rate-dependent influences on tissue response, causing either localized damage or slippage-based reorganization, based on the preconditioning level and loading rate. Based on these data, 10 cycles of 2\% preconditioning appear reasonable for tests conducted at 0.5\% strain per second, whereas tests conducted at 1\% strain per second may require higher preconditioning levels, 10\% strain or potentially even more, to obtain more stable responses. 

\subsubsection{Storage} 

The storage condition should also be considered as a protocol variable, with several studies showing that freeze-thaw cycles change material behavior \citep[]{mcgarvey_establishing_2025,jung_effects_2011}. Overall, freezing did not have a broad effect on the high-strain material behavior for all cases, suggesting that the high-strain, fiber-dominated behavior was largely unchanged. This is consistent with prior work showing that freezing can alter low-strain behavior while preserving overall stiffness \citep{leonard_effects_2022}, which may indicate that freezing affects the \gls{ecm} more than the embedded load-bearing fibers. 

Additionally, there is some evidence that pregnant tissue may be more affected by freezing. Linear models showed near significant effects of pregnancy and freezing interaction on both peak stress and stiffness.   This can also be seen in \Cref{fig:FigR_FreezingComparisonBoxplotPanel}.  Consider the results for both stiffness and peak stress in \Cref{fig:FigR_FreezingComparisonBoxplotPanel} at 10\% preconditioning.  The fresh control, the fresh pregnant, and the frozen control tissue have similar peak stress and stiffness, but there is a statistically significant or borderline statistically significant difference with the frozen pregnant case.  This suggests that the frozen pregnant tissue is different than the others.  
This is in line with literature that shows that freeze-thaw cycles can significantly change the high-strain material behavior in some soft tissues \citep[]{jung_effects_2011,leonard_effects_2022}. 
Moreover, the effect of pregnancy may be associated with water content.  In previous work, pregnancy is proven to increase the water content of lean tissue \citep{lactationBodyCompositionChanges1990}.  With this additional water and knowing that water expands as it freezes, the pregnant tissue likely experienced additional stretching during freezing, compared to the frozen control group.  This expansion could cause micro-damage to the tissue, which in turn could lead to overall damage even before uniaxial testing. And because the frozen pregnant tissue experienced lower peak stress and lower stiffness than the frozen control tissue or the fresh pregnant tissue, the micro-damage from freezing likely affected the high-strain behavior.

\subsection{Mechanical comparisons between pregnant and non-pregnant tissue}

Overall, pregnant rat pelvic floor samples generally exhibited greater deformation capacity than non-pregnant controls across the fresh-sample experimental comparisons. In the omnibus models, pregnancy condition  was a significant predictor of peak true strain, whereas peak true stress and high-strain linear stiffness did not show significant pregnancy effects. This suggests that pregnancy-related adaptation in the rat pelvic floor was expressed most clearly as increased extensibility rather than increased peak load-bearing capacity or bulk high strain-stiffness. This interpretation is supported bu the preconditioning comparison in \Cref{fig:FigR_PreconditionComparisonBoxplotPanel}, where pregnant samples generally reached higher peak true strains than non-pregnant controls across testing conditions. In contrast, peak stress and high-strain stiffness showed more variable subgroup-level behavior and were not supported by significant pregnancy effects in the corresponding omnibus models. Therefore, the increased strain capacity in pregnant samples should be interpreted as a deformation-capacity rather than evidence of uniformly greater strength or stiffness. 
The pooled literature comparison in \Cref{fig:FigR_LitComparison2} provides broader context for this finding. When fresh rat samples tested at 0.5\%s$^{-1}$ were pooled across all preconditioning levels, pregnant rat samples showed greater peak strains than control samples, though this trend was not significant, but reinforces the aggregate trend from the omnibus. 

Mechanistically, increased extensibility could arise from pregnancy-related remodeling of pelvic connective tissue and intramuscular \gls{ecm}. Prior work has shown that pregnancy and parturition are accompanied by biochemical and structural remodeling of the pelvic tissues, including changes in collagen organization and cross-linking \citep[]{alperin_pregnancy-induced_2015}. Such remodeling may allow greater inter-matrix sliding or reorganization during tensile loading, particularly before full fiber engagement. This mechanism would be consistent with increased peak strain without requiring corresponding increases in peak stress or high-strain stiffness. Structural models of passive muscle mechanics similarly support the idea that extracellular and macromolecular structures contribute to passive force transmission beyond normal actin--myosin filament overlap \citep[]{herzog_three_2012}.

The transition-region metrics did not show a clear pregnancy-condition effect, suggesting that the pregnancy-associated increase in peak strain was not primarily explained by a uniformly larger toe-region response. Rather, pregnancy-related adaptation may have emerged more clearly at larger strains, where passive load transfer through the intramuscular \gls{ecm}, fiber recruitment, and matrix--fiber interactions become more important. This interpretation is consistent with prior rat pelvic floor studies showing that pregnancy-induced adaptations attenuate sarcoma elongation during delivery-like vaginal distension, suggesting a protective response against hyper elongation-mediated muscle injury in some regions and decreased enzymatic cross-linking in others \citep[]{catanzarite_pelvic_2018}. Similarly, pregnancy has been shown to alter the intramuscular \gls{ecm} of pelvic floor muscles, increasing passive tension at longer muscle lengths while producing complex changes in collagen organization and cross-linking \citep[]{alperin_pregnancy-induced_2015}. Although pregnancy-associated changes in collagen cross-linking are complex, including decreased enzymatic cross-links and increased glycoside cross links, prior work suggests that these biochemical changes do not map straightforwardly onto a single stiffness metric \citep[]{alperin_pregnancy-induced_2015}. Therefore, the present finding of increased peak strain without a corresponding model-supported pregnancy effect on toe-region slope suggests that pregnancy-related remodeling was most clearly expressed in large-deformation tensile behavior in this dataset. However, because the toe region was variable and difficult to define consistently, these results do not exclude subtler pregnancy associated changes in the low-strain response.

\subsection{Nonlinearities in sample behavior and failure trends}

The pelvic floor muscle stress-strain curves did not typically exhibit a single clean linear region that could be selected unambiguously across samples using the traditional offset strain method employed for metals and other engineered materials. Instead, both the toe region and 
the high-strain ``linear region'' often showed piecewise behavior, local plateaus, sudden stress changes, and step-like or wave-like features.  
Irregular sample thickness and width through the gauge section contributed to high variability in the stress-strain response of each tissue, as shown in \Cref{fig:FigS_RepresentativeCurveWithUncertainty}. Although stress-based metrics were normalized by initial cross-sectional area, sensitivity analyses identified modest inverse associations between specifmen area and high-strain stiffness and peak stress. However, after adjustment for the experimental factors, area remained associated with peak stress but did not alter the interpretation of the primary experimental effects.
Rather than forming a consistently smooth and sharply bounded region, the early stress--strain response often transitioned gradually into the higher-stiffness loading regime. As a result, automated transition-point detection could be influenced by high signal-to-noise ratio, residual slack removal, and local curvature.  Moreover, because the toe region was defined as the region with the highest $R^2$ value for a linear fit in the early part of the stress--strain curve, when, as was the case for several samples, fits produced comparable $R^2$ values at several locations, the transition point may not be a clearly identifiable single point. 
This ambiguity was most apparent when the toe region was shorter or difficult to resolve, including some faster-rate tests. For this reason, transition stress, strain, and toe-region slope are best interpreted as repeatable operational metrics describing early-curve behavior, rather than precise material transition thresholds, and are similar to other protocols used to detect linear region stiffness metrics \citep[cf., e.g.,][]{khair_achilles_2024, noauthor_tendon_nodate}.

In the high-strain regime, \Cref{fig:FigS_MiniFractures} shows some of the sudden dips and plateaus present prior to failure.  
These features suggest that the pre-peak tensile response is not governed by a single uniform elastic regime, but by progressive loading coupled with localized small-scale structural disruption. 
As such, the stiffness metric used in this study should be interpreted as a repeatable measure of the bulk linear response maintained over sustained loading intervals, rather than as a material constant describing the entire stress-strain response.

\Cref{fig:FigS_MiniFractures} also shows that stepwise increases or changes in stress seem to correspond to visible local damage events that appeared before overall bulk failure.
These observations may also indicate that localized damage may accrue well before bulk failure.
Work by \citet{chittajallu_review_2022} discusses differences in energy dissipation, with a `softening' attributed to pre-failure slippage and localized fracture that may explain the irregularities in peak stress trends, which work by \citet{ateshian_damage_2022} suggests should increase as strain rate increases. 
Localized fracture, especially in passive anisotropic pelvic floor muscles, may differ from bulk failure, suggesting that fracture criteria for \glspl{pfm} may need to take into consideration local rather than purely global stress--strain behavior. 


\subsection{Comparisons to human pelvic floor muscles}

\begin{figure}[htbp!]
    \includegraphics[width=0.6\textwidth]{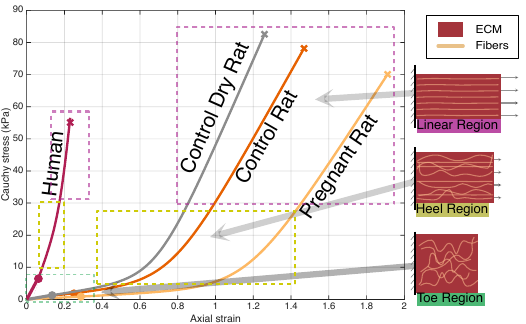}
    \caption{Idealized representative comparison of human (red) and non-pregnant control (orange), pregnant (yellow) and unsubmerged control (gray) rat mechanical behavior, with the transition point (depicted by a dot marker) and failure point (depicted by an `x' marker), and representative depictions of the relative fiber alignment in each phase.}
    \label{fig:FigS_LitCompDiagram}
\end{figure}

The comparison with human cadaveric \gls{lam} data provided important context for interpreting the translational relevance of the rat pelvic floor muscle model. The difference in strain rate likely explains the much higher peak stress and stiffness observed in the \citet{nagle_passive_2014} results. As \citet{nagle_passive_2014} used much higher strain rates than this work or the work of \citet{routzong_passive_2026}, the tissue had less time to relax, so stress is higher at a given strain, which leads to larger peak stress and stiffness.  This rate dependence of stiffness and peak stress has been observed in other biological materials \citep[e.g.,][]{ateshian_damage_2022}.   

The large deformation behavior of rat versus human \glspl{pfm} painted a stark image, which is captured in \Cref{fig:FigS_LitCompDiagram}. The most consistent differences between the mechanics of human cadaver and rat \gls{lam} were the peak strain and stiffness, where rat pelvic floor muscle samples reached substantially larger strains before failure than the digitized human \gls{lam} datasets, while having lower stiffnesses. This suggests that rat tissue may accommodate larger deformations than the non-pregnant human cadaver data, and human tissue is more resistant to deformation, although this difference may be attributed to other factors, such as tissue age, hydration during testing, storage methods, and other conditions. 
Most notably, hydration seems to be a likely candidate to explain this difference, as hydration is known to lubricate slippage at the molecular level, which leads to higher strains at the macroscopic level \citep[e.g.,][]{safa_exposure_2017, nicolle_dehydration_2010,abraham_phenomenological_2013}.  
This is further supported by the trend in the unsubmerged (dry) rat samples.  While the dry samples were kept moistened by an initial spray of \gls{pbs}, results for the dry samples seemed to bridge the gap between submerged rat and unsubmerged human samples in all metrics. It is also possible that the smaller size of the rat muscles leads to less fiber entanglement, and/or the smaller size of the rat muscles allows the fibers to relax more when removed from the body (in larger tissues, the neighboring material may prevent this additional relaxation), either of which could lead to larger peak strains. In contrast, peak stress showed more overlap between the rat and the \citet{routzong_passive_2026} human datasets, suggesting strength may be a cross-species material property, but further testing would be needed to confirm this.  

Toe metrics also show differences between the tissues. Human cadaver tissue from \citet{routzong_passive_2026} showed significantly higher transition stresses than both pregnant and control rat specimens, while exhibiting substantially lower transition strains, suggesting that the fibers may begin to engage significantly sooner in the tests on human samples than those on rat tissue. 
The toe stiffness was also higher in humans than in submerged rats, suggesting differences at the matrix level of the tissue. 
These matrix differences could be due to differences in hydration during testing, since hydration could lead to more lubrication and slippage prior to fiber engagement, as evidenced by the non-significant increase in dry rat toe-region stiffness.  This slippage would lead to higher strain and lower stiffness in the rat \glspl{pfm}. 
Overall, unsubmerged rat samples exhibited mechanical responses intermediate in magnitude between the submerged rat and human cadaveric datasets, and were not significantly different from either group for several metrics. In contrast, submerged rat and human samples different significantly in several metrics. This suggests that hydration may in part explain the differences observed in rat and human tissue; however, species dependence seems to be a larger factor. Additionally, hydration can alter tissue dimensions, frictional interactions, and apparent mechanical response \citep[]{hatami-marbini_effect_2025,wale_applying_2021,wang_hydration_2026}, but the direction and magnitude of these effects cannot be isolated in the present comparison because hydration state was not independently varied.

Despite these limitations, the pooled comparison suggests that rat pelvic floor muscle can reach comparable peak stresses and stiffness values within the same order of magnitude when tested under broadly comparable loading rates as human cadaveric \gls{lam}. However, the rat tissue undergoes substantially larger strains and has significantly different toe-region behavior as compared to human tissue. This indicates that the rat and human datasets may differ not only in magnitude, but also in the pathway through which the stress--strain response develops.  
Previous work \citep[]{binder-markey_systematic_2021}. \citet{binder-markey_systematic_2021} specifically discusses differences between fiber, fascicle, and bulk behavior between small and large laboratory mammals, which is in line with the data shown here.  
These comparisons should be interpreted cautiously because the testing environment may substantially affect measured tissue mechanics. Future studies should therefore report and control hydration conditions, including submersion duration and bath composition, when comparing pelvic floor muscle mechanics across datasets. 



\section{Conclusions}

To the best of our knowledge, this study is among the first to directly compare a host of mechanical properties of nulliparous non-pregnant and pregnant pelvic floor \glspl{lam} across pregnancy condition, preconditioning level, loading rate, and storage condition. Additionally, this study establishes a repeatable approach for quantifying bulk linear stiffness and transition stress/strain from nonlinear stress-strain responses. The results demonstrate that pregnant samples tended to reach higher peak strains than non-pregnant controls, suggesting that pregnancy-associated adaptation increased deformation capacity prior to bulk failure, without producing a corresponding increase in peak stress or high-strain stiffness. The protocol sweeps further demonstrated that the measured pelvic floor muscle behavior depends strongly on the test conditions imposed prior to failure. Preconditioning level and loading rate affected the observed response, supporting the need to control and report these parameters when evaluating passive pelvic floor muscle mechanics. 

The limitations in this study include limited sample sizes, exploratory protocol groups, and experimental differences between the present rat experiments and digitized human cadaver \gls{lam} datasets from the literature. Future work should pair mechanical testing with image-based strain and damage tracking, consider hydration as another important experimental protocol,  expand the sample sizes, and use these data for constitutive model calibration of large deformation pelvic floor mechanics. Despite these limitations, the present study demonstrates that the rat pelvic floor provides a useful experimental model for investigating pregnancy-associated mechanical adaptation, protocol sensitivity, and passive failure behavior, while also highlighting some potentially important differences between rat and human pelvic floor mechanics that should be considered when translating findings across species. Overall, these results indicate that pregnancy-associated deformation capacity and experimental testing conditions must both be considered when designing passive failure tests of pelvic floor muscle, interpreting results from these tests, and translating data from these tests into constitutive models. Future studies should extend these findings by incorporating localized stress--strain measurements to capture the spatial variations in mechanical response observed in \Cref{fig:FigS_MiniFractures}, as the heterogeneous structure of skeletal muscle can produce substantial variations in local stress and strain \citep[cf.][]{mcguire_tear_2021,azizi_regional_2014}. Spatial variation and uncertainty in specimen cross-sectional area should likewise be characterized when possible, particularly given the geometric variability observed in \Cref{fig:FigS_RepresentativeCurveWithUncertainty}. Hydration conditions should also be controlled and reported consistently because hydration protocols can alter measured properties \citep[]{safa_exposure_2017,wang_hydration_2026,hatami-marbini_effect_2025}. Together, these improvements would support more reproducible characterization of passive \gls{pfm} mechanics and the development of constitutive models that account for pregnancy-associated remodeling and rate-dependent behavior, ultimately improving simulation of pelvic floor deformation and injury during childbirth. 
\section*{Acknowledgments} This material is based upon work supported by the National Science Foundation under Grant
No. 2349258. We thank Kylie Akey for her contributions to the manuscript.
\section*{Supplemental Materials}

\setcounter{figure}{0}
\renewcommand{\thefigure}{S\arabic{figure}}
\setcounter{table}{0} 
\renewcommand{\thetable}{S\arabic{table}}
\renewcommand{\thesubsection}{S.\arabic{subsection}}

\subsection{Data used in box plot comparisons}
\label{s:data_in_boxplots}

\begin{figure}[htbp!]
    \centering
    \includegraphics[width=1\textwidth]{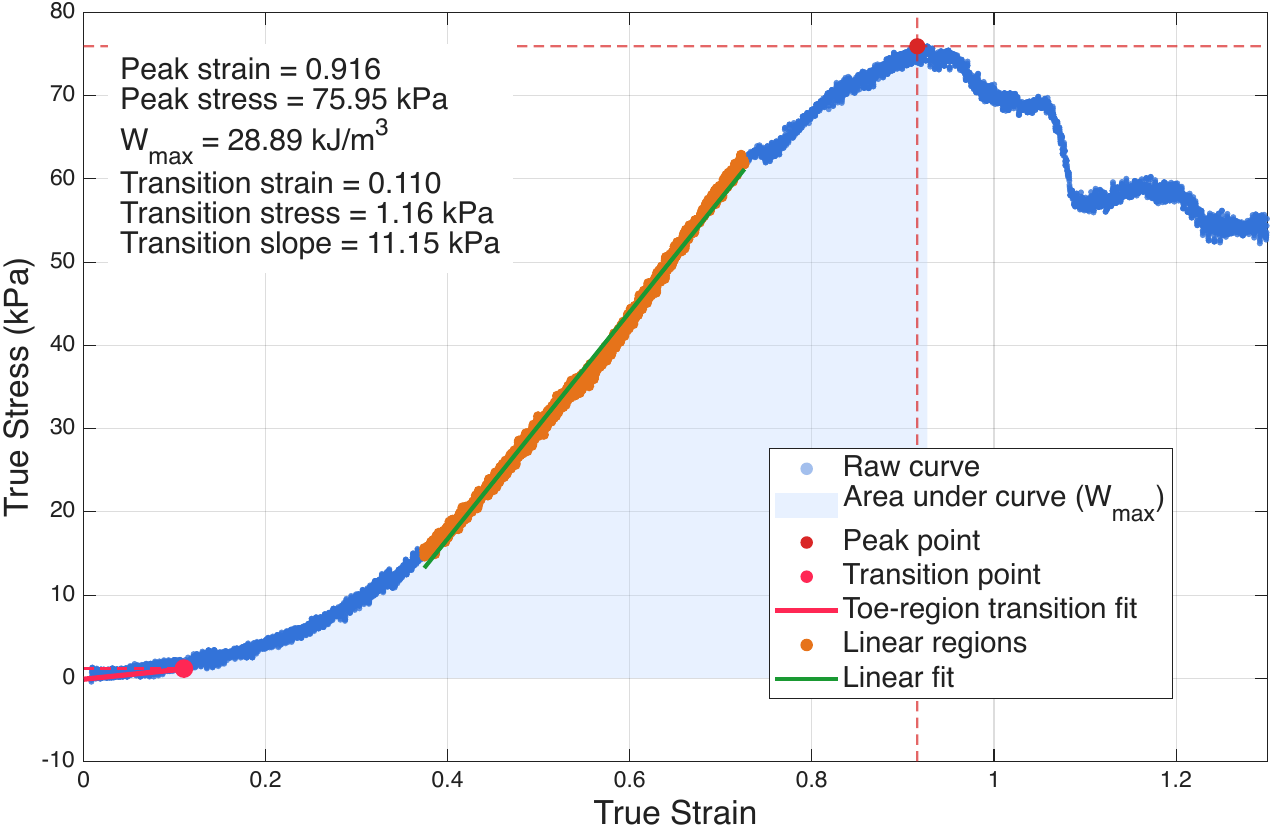}
    \caption{Representative raw stress versus true strain curve with peak stress and peak strain marked with a red dot, the detected stiffness of the linear region of the tissue highlighted in yellow, with green lines showing the fit line for that region, and the $W_{max}$ values calculated as the area under the curve for pre-peak data (shaded blue region).}
    \label{fig:FigS_PeakValuesDemo}
\end{figure}

\Cref{fig:FigS_PeakValuesDemo} depicts a representative raw stress-strain curves used to compare tissue behavior across experimental conditions, including peak stress, peak strain, pre-peak strain energy density ($W_{max}$), and bulk linear stiffness. Automated detection of these metrics allowed for repeatable comparisons across all experimental conditions. 

\Cref{fig:FigS_SavitskyGolayPanel} depicts the six step process for finding the average linear region stiffness regardless of sample noise or small scale stepping as a result of localized fracture. 
First, the pre-peak stress-strain curve was temporarily smoothed with a Savitsky--Golay filter for detection purposes only. A centered rolling-window linear regressions was used to estimate local tangent modulus and local $R^2$ values. 
Then, candidate loading regions were identified using a two-stage detection. First, the early toe region was excluded using an adaptive search. Points were only considered after the curve exceeded 30\% of the total strain range and 10\% of the maximum stress. Within this search domain, centered rolling-window linear regressions were applied to the temporarily smoothed stress-strain curve using short and long strain windows of 0.03 and 0.10 strain, respectively. Candidate points were required to have a positive slope and a long window with $R^2 > 0.85$. Then, candidate points were  grouped into continuous interval chunks, with short chunk 'islands' removed and small gaps (smaller than 0.01 strain) bridged. Next, each was locally trimmed to retain the sustained region near its peak long-window slope, and chunks were retained as eligible regions only if they were within 25\% of the global peak long-window slope, had a raw data linear fit with $R^2 \geq 0.90$, positive stress rise, and a non-zero raw fit stress. Lastly, the mean slope across all eligible chunks was recorded as the bulk linear region stiffness.

\Cref{fig:S03_raw-slope-curves} shows the raw data for each sample in the study along with their detected linear region stiffness overlayed on the raw curve. The samples show a robust methodology used to detect linear region stiffness regardless of the stepping and plateauing artifacts common in anisotropic soft tissues, allowing for repeatable comparison metrics. 
%


\begin{figure}[htbp!]
    \centering
    \includegraphics[width=1\textwidth]{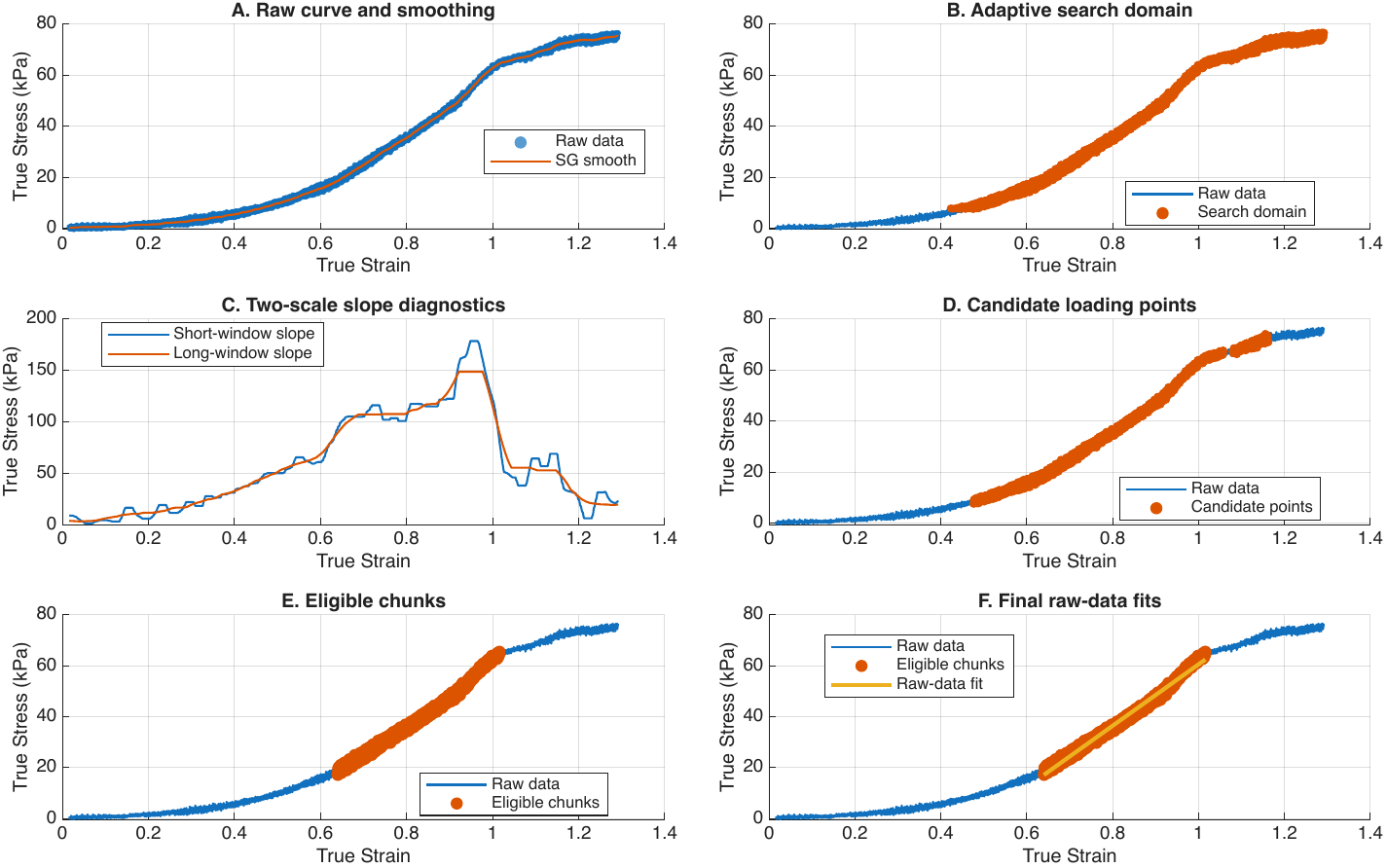}
    \caption{Step-wise detection of the bulk linear response in noisy pelvic floor muscle stress-strain curves. 
    A. Raw stress-strain data were temporarily smoothed using a Savitzky--Golay filter to reduce high-frequency noise while preserving the overall curve shape. 
    B. An adaptive search domain was defined to exclude the early toe-region response. Candidate points were only considered after the curve exceeded 30\% of the total strain range and 10\% of the maximum stress. 
    C. Local tangent stiffness was estimated using centered rolling-window linear regressions at two strain-window scales (fine/short window, and coarse/long window), allowing the detector to identify sustained regions of elevated slope and local linearity. 
    D. Candidate loading points were selected based on local slope, stress level, and linearity criteria. 
    E. Candidate points were grouped into continuous chunks, and eligible chunks were retained based on strain span, stress rise, and raw-data fit quality, with the highest slope chosen as the linear region slope, due to assumptions of Fung-type anisotropic behavior \citep[]{de_pascalis_nonlinear_2014}. 
    F. Linear polynomials were fit to the corresponding raw, un-smoothed stress-strain data within the eligible chunks. The resulting slope was used as the bulk linear stiffness metric for each sample.}
    \label{fig:FigS_SavitskyGolayPanel}
\end{figure}









\newcounter{RawSlopePage}

\foreach \page in {01,02,03,04,05,06}{%
    \begin{figure}[p]

        \ifnum\value{RawSlopePage}>0
            \ContinuedFloat
        \fi

        \centering

        \begin{subfigure}{\textwidth}
            \centering
            \ifdefstring{\page}{07}{%
                \includegraphics[\textwidth]{Ch99_SupplementaryInfo/Supp_Figures/Supp_RawSlopeCurves/FigS_LinearRegionDetectorTiles_Page\page.pdf}%
            }{%
                \includegraphics[width=\textwidth]{Ch99_SupplementaryInfo/Supp_Figures/Supp_RawSlopeCurves/FigS_LinearRegionDetectorTiles_Page\page.pdf}%
            }
            \phantomsubcaption
        \end{subfigure}

        \ifnum\value{RawSlopePage}=0
            \caption{Final linear-region intervals identified for the
            true stress--strain curves of all tested datasets.}
            \label{fig:S03_raw-slope-curves}
        \else
            \caption[]{Final linear-region intervals identified for the
            true stress--strain curves of all tested datasets, continued.}
        \fi

    \end{figure}

    \stepcounter{RawSlopePage}
}











\FloatBarrier

\subsection{Summary of all experimental conditions and details on each rat and sample}

\Cref{tab:TableS01_ExperimentalConditionSummary} details all the experiments performed, including the pregnancy condition, storage condition, loading rate, preconditioning strain amplitude, initial cross sectional area ($A_0$), variability in cross-sectional area ($dA$) as calculating using uncertainty propagation and the standard deviation that arose from different measurements of thickness and width, and the mass of each rat prior to dissection.  

\begin{landscape}
\begin{table}[ht]
\caption{Summary of all rat specimens used in experiential comparisons for all conditions, with included pregnancy condition, tissue storage condition, preconditioning strain amplitude, average area measurement, variation in the area measurement, mass of the rat before tissue excision, along with the eight metrics presented in the boxplots.}
\label{tab:TableS01_ExperimentalConditionSummary}
\resizebox{1.4\textwidth}{!}{
\begin{tabular}{l|llllllllllllllll}
RatID & Condition & \makecell{TrueStress\\(kPa)} & Submersion & \makecell{Rate\\(\%/s)} & \makecell{Precon\\(\%)} & \makecell{A$_0$\\(mm$^2$)} & \makecell{dA\\(mm$^2$)} & \makecell{mass\\(g)} & \makecell{TrueStress\\(kPa)} & Strain & \makecell{Stiffness\\(kPa)} & \makecell{W$_{max}$\\($\frac{kJ}{m^3}$)} & \makecell{Transition\\Stress (kPa)} & \makecell{Transition\\Strain} & \makecell{Transition\\Slope (kPa)} & \makecell{Precon\\Cycle 10\\Area ($\frac{kJ}{m^3}$)} \\ \hline
R02\_Control\_Left\_Pre05\_Frozen\_Rate1p0\_Submerged\_250806.xlsx   & Control   & Frozen       & Submerged   & 1           & 5           & 15.19                        & 3.54                       & 0.00     & 98.12             & 0.98   & 143.20          & 45.47                          & 4.40                    & 0.13              & 34.49                  & 0.01                                           \\
R02\_Control\_Right\_Pre10\_Frozen\_Rate1p0\_Submerged\_250806.xlsx  & Control   & Frozen       & Submerged   & 1           & 10          & 16.49                        & 3.73                       & 0.00     & 105.80            & 1.17   & 146.47          & 44.13                          & 1.42                    & 0.13              & 11.09                  & 0.01                                           \\
R03\_Pregnant\_Left\_Pre05\_Frozen\_Rate1p0\_Submerged\_250806.xlsx  & Pregnant  & Frozen       & Submerged   & 1           & 5           & 17.94                        & 5.24                       & 349.76   & 103.15            & 1.50   & 109.69          & 70.60                          & 4.72                    & 0.16              & 30.18                  & 0.00                                           \\
R03\_Pregnant\_Right\_Pre10\_Frozen\_Rate1p0\_Submerged\_250806.xlsx & Pregnant  & Frozen       & Submerged   & 1           & 10          & 29.83                        & 9.14                       & 349.76   & 49.45             & 1.03   & 93.69           & 26.20                          & 12.20                   & 0.39              & 32.25                  & 0.01                                           \\
R04\_Control\_Left\_Pre05\_Frozen\_Rate1p0\_Submerged\_250806.xlsx   & Control   & Frozen       & Submerged   & 1           & 5           & 18.34                        & 8.66                       & 177.98   & 76.56             & 1.29   & 120.58          & 37.96                          & 2.57                    & 0.32              & 7.52                   & 0.00                                           \\
R04\_Control\_Right\_Pre10\_Frozen\_Rate1p0\_Submerged\_250806.xlsx  & Control   & Frozen       & Submerged   & 1           & 10          & 29.36                        & 5.85                       & 177.98   & 118.87            & 1.22   & 182.47          & 51.43                          & 3.52                    & 0.29              & 12.76                  & 0.00                                           \\
R05\_Pregnant\_Left\_Pre05\_Frozen\_Rate1p0\_Submerged\_250806.xlsx  & Pregnant  & Frozen       & Submerged   & 1           & 5           & 44.56                        & 3.64                       & 335.03   & 26.96             & 0.94   & 55.68           & 9.19                           & 0.45                    & 0.15              & 3.18                   & 0.00                                           \\
R05\_Pregnant\_Right\_Pre10\_Frozen\_Rate1p0\_Submerged\_250806.xlsx & Pregnant  & Frozen       & Submerged   & 1           & 10          & 44.56                        & 3.28                       & 335.03   & 53.81             & 2.44   & 48.49           & 44.08                          & 1.09                    & 0.28              & 4.07                   & 0.00                                           \\
R06\_Pregnant\_Left\_Pre15\_Frozen\_Rate1p0\_Submerged\_250806.xlsx  & Pregnant  & Frozen       & Submerged   & 1           & 15          & 32.61                        & 4.23                       & 376.08   & 51.42             & 1.99   & 50.93           & 40.42                          & 6.29                    & 0.75              & 8.60                   & 0.01                                           \\
R06\_Pregnant\_Right\_Pre15\_Frozen\_Rate1p0\_Submerged\_250806.xlsx & Pregnant  & Frozen       & Submerged   & 1           & 15          & 16.52                        & 1.81                       & 376.08   & 35.81             & 1.82   & 57.91           & 26.86                          & 2.42                    & 0.34              & 7.16                   & 0.00                                           \\
R07\_Control\_Left\_Pre10\_Fresh\_Rate1p0\_Submerged\_251118.csv     & Control   & Fresh        & Submerged   & 1           & 10          & 19.01                        & 5.58                       & 257.66   & 101.28            & 1.45   & 114.84          & 59.08                          & 3.13                    & 0.15              & 21.16                  & 0.02                                           \\
R07\_Control\_Right\_Pre00\_Fresh\_Rate1p0\_Submerged\_251118.csv    & Control   & Fresh        & Submerged   & 1           & 0           & 12.82                        & 2.93                       & 257.66   & 34.51             & 1.10   & 49.82           & 21.23                          & 2.21                    & 0.10              & 22.27                  & NaN                                            \\
R08\_Control\_Left\_Pre00\_Fresh\_Rate1p0\_Submerged\_251118.csv     & Control   & Fresh        & Submerged   & 1           & 0           & 8.69                         & 3.78                       & 239.92   & 50.79             & 1.66   & 45.71           & 36.43                          & 6.23                    & 0.23              & 26.55                  & NaN                                            \\
R09\_Pregnant\_Left\_Pre10\_Fresh\_Rate1p0\_Submerged\_251118.csv    & Pregnant  & Fresh        & Submerged   & 1           & 10          & 21.53                        & 6.24                       & 337.34   & 91.06             & 0.89   & 167.70          & 29.26                          & 10.90                   & 0.31              & 44.05                  & 0.01                                           \\
R09\_Pregnant\_Right\_Pre00\_Fresh\_Rate1p0\_Submerged\_251118.csv   & Pregnant  & Fresh        & Submerged   & 1           & 0           & 21.05                        & 7.67                       & 337.34   & 81.65             & 1.18   & 101.79          & 37.42                          & 0.75                    & 0.15              & 4.78                   & NaN                                            \\
R10\_Pregnant\_Left\_Pre00\_Fresh\_Rate1p0\_Submerged\_251118.csv    & Pregnant  & Fresh        & Submerged   & 1           & 0           & 30.27                        & 11.42                      & 314.05   & 45.86             & 1.29   & 70.18           & 27.27                          & 4.74                    & 0.33              & 14.62                  & NaN                                            \\
R11\_Control\_Right\_Pre05\_Fresh\_Rate0p5\_Submerged\_260324.csv    & Control   & Fresh        & Submerged   & 0.5         & 5           & 24.03                        & 4.78                       & 220.33   & 2.34              & 0.25   & NaN             & 0.21                           & 0.67                    & 0.14              & 4.73                   & 0.00                                           \\
R12\_Pregnant\_Left\_Pre02\_Fresh\_Rate0p5\_Submerged\_260324.csv    & Pregnant  & Fresh        & Submerged   & 0.5         & 2           & 20.03                        & 3.40                       & 287.21   & 63.11             & 1.33   & 101.85          & 30.45                          & 1.94                    & 0.29              & 6.56                   & 0.00                                           \\
R13\_Control\_Right\_Pre10\_Fresh\_Rate0p5\_Submerged\_260324.csv    & Control   & Fresh        & Submerged   & 0.5         & 10          & 20.38                        & 4.00                       & 199.02   & 34.16             & 1.31   & 49.27           & 16.57                          & 3.45                    & 0.40              & 8.87                   & 0.01                                           \\
R14\_Pregnant\_Left\_Pre10\_Fresh\_Rate0p5\_Submerged\_260325.csv    & Pregnant  & Fresh        & Submerged   & 0.5         & 10          & 44.59                        & 8.09                       & 305.63   & 65.58             & 1.75   & 86.06           & 39.63                          & 0.50                    & 0.19              & 2.66                   & 0.00                                           \\
R14\_Pregnant\_Right\_Pre02\_Fresh\_Rate0p5\_Submerged\_260325.csv   & Pregnant  & Fresh        & Submerged   & 0.5         & 2           & 17.39                        & 3.39                       & 305.63   & 113.02            & 1.89   & 153.93          & 64.88                          & 0.74                    & 0.29              & 2.68                   & 0.00                                           \\
R15\_Control\_Left\_Pre02\_Fresh\_Rate0p5\_Submerged\_260325.csv     & Control   & Fresh        & Submerged   & 0.5         & 2           & 15.18                        & 2.36                       & 200.72   & 75.95             & 0.92   & 136.25          & 28.89                          & 1.16                    & 0.11              & 11.15                  & 0.00                                           \\
R15\_Control\_Right\_Pre00\_Fresh\_Rate0p5\_Submerged\_260325.csv    & Control   & Fresh        & Submerged   & 0.5         & 0           & 23.66                        & 5.36                       & 200.72   & 162.98            & 1.86   & 213.29          & 89.24                          & 0.47                    & 0.23              & 1.90                   & NaN                                            \\
R16\_Pregnant\_Left\_Pre00\_Fresh\_Rate0p5\_Submerged\_260325.csv    & Pregnant  & Fresh        & Submerged   & 0.5         & 0           & 24.45                        & 7.33                       & 294.80   & 29.97             & 1.24   & 50.59           & 13.94                          & 0.40                    & 0.14              & 3.52                   & NaN                                            \\
R16\_Pregnant\_Right\_Pre02\_Fresh\_Rate0p5\_Submerged\_260325.csv   & Pregnant  & Fresh        & Submerged   & 0.5         & 2           & 20.44                        & 4.30                       & 294.80   & 159.84            & 1.56   & 250.77          & 78.19                          & 0.65                    & 0.23              & 2.60                   & 0.00                                           \\
R17\_Control\_Right\_Pre15\_Fresh\_Rate0p5\_Submerged\_260325.csv    & Control   & Fresh        & Submerged   & 0.5         & 15          & 22.37                        & 4.50                       & 202.62   & 64.74             & 1.30   & 87.25           & 33.40                          & 9.37                    & 0.60              & 16.14                  & 0.01                                           \\
R18\_Control\_Left\_Pre00\_Fresh\_Rate0p5\_Submerged\_260512.csv     & Control   & Fresh        & Submerged   & 0.5         & 0           & 22.67                        & 6.03                       & 193.44   & 108.78            & 1.80   & 128.61          & 64.08                          & 0.46                    & 0.13              & 3.26                   & NaN                                            \\
R18\_Control\_Right\_Pre05\_Fresh\_Rate0p5\_Submerged\_260512.csv    & Control   & Fresh        & Submerged   & 0.5         & 5           & 21.14                        & 4.69                       & 193.44   & 51.51             & 1.27   & 90.85           & 20.89                          & 1.44                    & 0.21              & 6.88                   & 0.00                                           \\
R19\_Pregnant\_Left\_Pre05\_Fresh\_Rate0p5\_Submerged\_260512.csv    & Pregnant  & Fresh        & Submerged   & 0.5         & 5           & 32.81                        & 12.22                      & 283.84   & 33.81             & 1.12   & 68.90           & 12.36                          & 0.29                    & 0.12              & 2.48                   & 0.00                                           \\
R19\_Pregnant\_Right\_Pre15\_Fresh\_Rate0p5\_Submerged\_260512.csv   & Pregnant  & Fresh        & Submerged   & 0.5         & 15          & 28.24                        & 13.36                      & 283.84   & 54.31             & 2.97   & 56.66           & 87.84                          & 0.75                    & 0.24              & 3.11                   & 0.00                                           \\
R20\_Pregnant\_Left\_Pre15\_Fresh\_Rate0p5\_Submerged\_260512.csv    & Pregnant  & Fresh        & Submerged   & 0.5         & 15          & 22.09                        & 6.65                       & 321.28   & 59.58             & 1.76   & 78.25           & 32.15                          & 1.42                    & 0.58              & 2.49                   & 0.01                                           \\
R20\_Pregnant\_Right\_Pre05\_Fresh\_Rate0p5\_Submerged\_260512.csv   & Pregnant  & Fresh        & Submerged   & 0.5         & 5           & 39.80                        & 15.78                      & 321.28   & 36.57             & 2.82   & 33.01           & 50.84                          & 1.35                    & 0.37              & 3.78                   & 0.00                                           \\
R21\_Control\_Left\_Pre10\_Fresh\_Rate0p5\_Submerged\_260513.csv     & Control   & Fresh        & Submerged   & 0.5         & 10          & 31.64                        & 6.26                       & 159.59   & 35.03             & 1.36   & 59.99           & 17.21                          & 2.09                    & 0.35              & 6.34                   & 0.00                                           \\
R21\_Control\_Right\_Pre15\_Fresh\_Rate0p5\_Submerged\_260513.csv    & Control   & Fresh        & Submerged   & 0.5         & 15          & 18.85                        & 13.22                      & 159.59   & 134.68            & 1.67   & 164.83          & 101.37                         & 0.30                    & 0.12              & 2.39                   & 0.00                                           \\
R22\_Control\_Left\_Pre15\_Fresh\_Rate0p5\_Submerged\_260513.csv     & Control   & Fresh        & Submerged   & 0.5         & 15          & 28.09                        & 5.24                       & 192.83   & 34.45             & 1.40   & 49.14           & 16.99                          & 1.65                    & 0.38              & 3.13                   & 0.01                                           \\
R22\_Control\_Right\_Pre10\_Fresh\_Rate0p5\_Submerged\_260513.csv    & Control   & Fresh        & Submerged   & 0.5         & 10          & 18.89                        & 4.26                       & 192.83   & 48.88             & 1.68   & 61.06           & 28.33                          & 2.06                    & 0.40              & 5.24                   & 0.00                                           \\
R23\_Pregnant\_Left\_Pre00\_Fresh\_Rate0p5\_Submerged\_260513.csv    & Pregnant  & Fresh        & Submerged   & 0.5         & 0           & 22.93                        & 6.95                       & 270.14   & 107.17            & 2.83   & 79.42           & 115.36                         & 0.49                    & 0.32              & 1.50                   & NaN                                            \\
R24\_Pregnant\_Left\_Pre10\_Fresh\_Rate0p5\_Submerged\_260513.csv    & Pregnant  & Fresh        & Submerged   & 0.5         & 10          & 15.84                        & 12.39                      & 303.17   & 59.68             & 2.35   & 58.63           & 49.89                          & 0.48                    & 0.55              & 0.70                   & 0.00                                           \\
R24\_Pregnant\_Right\_Pre15\_Fresh\_Rate0p5\_Submerged\_260513.csv   & Pregnant  & Fresh        & Submerged   & 0.5         & 15          & 22.03                        & 5.21                       & 303.17   & 13.54             & 2.15   & 13.35           & 8.16                           & 0.13                    & 0.24              & 0.67                   & 0.00                                           \\
R25\_Control\_Left\_Pre05\_Fresh\_Rate0p5\_Submerged\_260514.csv     & Control   & Fresh        & Submerged   & 0.5         & 5           & 24.01                        & 6.96                       & 198.46   & 158.92            & 2.01   & 128.70          & 149.08                         & 0.30                    & 0.06              & 5.46                   & 0.00                                           \\
R25\_Control\_Right\_Pre02\_Fresh\_Rate0p5\_Submerged\_260514.csv    & Control   & Fresh        & Submerged   & 0.5         & 2           & 23.04                        & 10.82                      & 198.46   & 79.96             & 1.34   & 116.16          & 42.34                          & 1.32                    & 0.23              & 5.74                   & 0.00                                           \\
R26\_Control\_Left\_Pre00\_Fresh\_Rate0p5\_Submerged\_260514.csv     & Control   & Fresh        & Submerged   & 0.5         & 0           & 18.87                        & 5.56                       & 203.63   & 33.80             & 0.93   & 63.49           & 10.43                          & 1.32                    & 0.18              & 7.28                   & NaN                                            \\
R26\_Control\_Right\_Pre00\_Fresh\_Rate1p0\_Submerged\_260514.csv    & Control   & Fresh        & Submerged   & 1           & 0           & 23.69                        & 5.32                       & 203.63   & 36.53             & 0.75   & 74.07           & 8.44                           & 0.37                    & 0.07              & 6.07                   & NaN                                            \\
R27\_Pregnant\_Right\_Pre10\_Fresh\_Rate0p5\_Submerged\_260514.csv   & Pregnant  & Fresh        & Submerged   & 0.5         & 10          & 13.31                        & 3.69                       & 300.24   & 155.96            & 1.61   & 212.25          & 75.55                          & 3.37                    & 0.56              & 5.54                   & 0.01                                           \\
R28\_Pregnant\_Left\_Pre10\_Fresh\_Rate1p0\_Submerged\_260514.csv    & Pregnant  & Fresh        & Submerged   & 1           & 10          & 12.44                        & 2.19                       & 282.09   & 129.81            & 2.04   & 111.19          & 92.26                          & 1.40                    & 0.27              & 5.03                   & 0.00                                           \\
R28\_Pregnant\_Right\_Pre00\_Fresh\_Rate0p5\_Submerged\_260514.csv   & Pregnant  & Fresh        & Submerged   & 0.5         & 0           & 27.18                        & 4.83                       & 282.09   & 42.67             & 1.99   & 37.47           & 38.58                          & 0.31                    & 0.10              & 3.13                   & NaN                                            \\
R29\_Pregnant\_Left\_Pre05\_Fresh\_Rate0p5\_Submerged\_260515.csv    & Pregnant  & Fresh        & Submerged   & 0.5         & 5           & 29.38                        & 7.71                       & 310.48   & 56.19             & 1.30   & 73.03           & 28.28                          & 0.57                    & 0.13              & 3.88                   & 0.00                                           \\
R29\_Pregnant\_Right\_Pre10\_Fresh\_Rate1p0\_Submerged\_260515.csv   & Pregnant  & Fresh        & Submerged   & 1           & 10          & 25.74                        & 6.51                       & 310.48   & 57.41             & 1.50   & 101.81          & 40.41                          & 0.45                    & 0.12              & 3.25                   & 0.00                                           \\
R30\_Control\_Left\_Pre10\_Fresh\_Rate1p0\_Submerged\_260515.csv     & Control   & Fresh        & Submerged   & 1           & 10          & 15.31                        & 4.48                       & 208.24   & 134.07            & 1.54   & 214.12          & 58.78                          & 1.15                    & 0.21              & 5.73                   & 0.00                                           \\
R30\_Control\_Right\_Pre02\_Fresh\_Rate0p5\_Submerged\_260515.csv    & Control   & Fresh        & Submerged   & 0.5         & 2           & 20.16                        & 5.68                       & 208.24   & 70.43             & 1.71   & 118.84          & 65.03                          & 0.39                    & 0.15              & 2.31                   & 0.00                                           \\
R31\_Control\_Right\_Pre10\_Fresh\_Rate1p0\_Submerged\_260515.csv    & Control   & Fresh        & Submerged   & 1           & 10          & 20.69                        & 5.07                       & 206.82   & 80.64             & 1.30   & 147.60          & 33.29                          & 0.74                    & 0.22              & 3.58                   & 0.00                                           \\
R32\_Control\_Right\_Pre02\_Fresh\_Rate0p5\_Unsubmerged\_260909.csv  & Control   & Fresh        & Unsubmerged & 0.5         & 2           & 18.48                        & 5.58                       & 182.81   & 113.76            & 1.12   & 198.46          & 55.00                          & 1.58                    & 0.14              & 10.36                  & 0.00                                           \\
R33\_Control\_Left\_Pre02\_Fresh\_Rate0p5\_Unsubmerged\_260909.csv   & Control   & Fresh        & Unsubmerged & 0.5         & 2           & 20.58                        & 4.98                       & 186.56   & 149.27            & 1.93   & 217.36          & 121.54                         & 1.92                    & 0.21              & 9.32                   & 0.00                                           \\
R33\_Control\_Right\_Pre02\_Fresh\_Rate0p5\_Unsubmerged\_260909.csv  & Control   & Fresh        & Unsubmerged & 0.5         & 2           & 38.56                        & 10.11                      & 186.56   & 42.92             & 1.07   & 72.49           & 17.91                          & 0.32                    & 0.08              & 3.81                   & 0.00                                           \\
R36\_Control\_Left\_Pre02\_Fresh\_Rate0p5\_Unsubmerged\_260910.csv   & Control   & Fresh        & Unsubmerged & 0.5         & 2           & 11.70                        & 6.43                       & 175.59   & 24.35             & 0.92   & 48.51           & 11.41                          & 1.39                    & 0.12              & 12.67                  & 0.00                                          
\end{tabular}%
}
\end{table}
\end{landscape}

\begin{landscape}
\begin{table}[ht]
\caption{Group-level mean $\pm$ standard deviation for specimen measurements and mechanical metrics across all experimental conditions. Groups are defined by pregnancy condition, tissue storage condition, submersion condition, loading rate, and preconditioning strain amplitude.}
\label{tab:TableS02_ExperimentalGroupSummary}
\resizebox{1.25\textwidth}{!}{%
\begin{tabular}{lllllr|lllllllllll}
\makecell{Tissue\\Condition} &
Storage &
Submersion &
\makecell{Rate\\(\%/s)} &
\makecell{Precon\\(\%)} &
$n$ &
\makecell{$A_0$\\(mm$^2$)} &
\makecell{Mass\\(g)} &
\makecell{$L_0$\\(mm)} &
\makecell{Peak Cauchy\\Stress (kPa)} &
\makecell{Peak Axial\\Strain} &
\makecell{High-Strain\\Stiffness (kPa)} &
\makecell{$W_{\max}$\\($\frac{\mathrm{kJ}}{\mathrm{m}^3}$)} &
\makecell{Transition\\Stress (kPa)} &
\makecell{Transition\\Strain} &
\makecell{Transition\\Slope (kPa)} &
\makecell{Precon Cycle 10\\Area ($\frac{\mathrm{kJ}}{\mathrm{m}^3}$)}
\\ \hline
Control & Fresh & Submerged  & 0.5 & 0 & 3 & 21.74 $\pm$ 2.53 & 199.26 $\pm$ 5.25 & 5.95 $\pm$ 0.07 & 101.85 $\pm$ 64.86 & 1.53 $\pm$ 0.52 & 135.13 $\pm$ 75.11 & 54.58 $\pm$ 40.25 & 0.75 $\pm$ 0.49 & 0.18 $\pm$ 0.05 & 4.15 $\pm$ 2.80 & -- \\
Control & Fresh & Unsubmerged & 0.5 & 2 & 4 & 30.89 $\pm$ 13.49 & 186.25 $\pm$ 2.58 & 5.68 $\pm$ 0.51 & 84.28 $\pm$ 56.64 & 1.36 $\pm$ 0.39 & 131.28 $\pm$ 90.01 & 55.55 $\pm$ 46.71 & 0.96 $\pm$ 0.93 & 0.13 $\pm$ 0.06 & 6.19 $\pm$ 4.36 & 0.0005 $\pm$ 0.0001 \\
Control & Fresh & Submerged  & 0.5 & 2 & 3 & 19.46 $\pm$ 3.97 & 202.47 $\pm$ 5.12 & 6.09 $\pm$ 0.41 & 75.45 $\pm$ 4.79 & 1.32 $\pm$ 0.40 & 123.75 $\pm$ 10.91 & 45.42 $\pm$ 18.27 & 0.96 $\pm$ 0.50 & 0.16 $\pm$ 0.06 & 6.40 $\pm$ 4.46 & 0.0007 $\pm$ 0.0001 \\
Control & Fresh & Submerged  & 0.5 & 5 & 3 & 23.06 $\pm$ 1.66 & 204.08 $\pm$ 14.30 & 5.88 $\pm$ 0.06 & 70.92 $\pm$ 80.08 & 1.18 $\pm$ 0.89 & 109.78 $\pm$ 26.76 & 56.73 $\pm$ 80.65 & 0.80 $\pm$ 0.58 & 0.13 $\pm$ 0.07 & 5.69 $\pm$ 1.09 & 0.0016 $\pm$ 0.0003 \\
Control & Fresh & Submerged  & 0.5 & 10 & 3 & 23.63 $\pm$ 6.97 & 183.81 $\pm$ 21.21 & 6.28 $\pm$ 0.62 & 39.36 $\pm$ 8.26 & 1.45 $\pm$ 0.20 & 56.77 $\pm$ 6.52 & 20.71 $\pm$ 6.61 & 2.53 $\pm$ 0.79 & 0.38 $\pm$ 0.03 & 6.82 $\pm$ 1.86 & 0.0037 $\pm$ 0.0028 \\
Control & Fresh & Submerged  & 0.5 & 15 & 3 & 23.10 $\pm$ 4.66 & 185.01 $\pm$ 22.55 & 6.72 $\pm$ 0.61 & 77.96 $\pm$ 51.40 & 1.46 $\pm$ 0.19 & 100.40 $\pm$ 58.96 & 50.59 $\pm$ 44.74 & 3.77 $\pm$ 4.89 & 0.36 $\pm$ 0.24 & 7.22 $\pm$ 7.74 & 0.0101 $\pm$ 0.0048 \\
Control & Fresh & Submerged  & 1 & 0 & 3 & 15.07 $\pm$ 7.75 & 233.74 $\pm$ 27.54 & 7.08 $\pm$ 0.64 & 40.61 $\pm$ 8.87 & 1.17 $\pm$ 0.46 & 56.53 $\pm$ 15.32 & 22.03 $\pm$ 14.01 & 2.94 $\pm$ 3.00 & 0.13 $\pm$ 0.09 & 18.30 $\pm$ 10.80 & -- \\
Control & Fresh & Submerged  & 1 & 10 & 3 & 18.33 $\pm$ 2.75 & 224.24 $\pm$ 28.95 & 6.26 $\pm$ 0.23 & 105.33 $\pm$ 26.94 & 1.43 $\pm$ 0.12 & 158.85 $\pm$ 50.59 & 50.38 $\pm$ 14.81 & 1.67 $\pm$ 1.28 & 0.19 $\pm$ 0.04 & 10.16 $\pm$ 9.59 & 0.0082 $\pm$ 0.0075 \\
Control & Frozen & Submerged  & 1 & 5 & 2 & 16.77 $\pm$ 2.23 & 88.99 $\pm$ 12.58 & 6.44 $\pm$ 0.32 & 87.34 $\pm$ 15.25 & 1.14 $\pm$ 0.21 & 131.89 $\pm$ 16.00 & 41.71 $\pm$ 5.31 & 3.48 $\pm$ 1.29 & 0.23 $\pm$ 0.13 & 21.01 $\pm$ 19.07 & 0.0036 $\pm$ 0.0032 \\
Control & Frozen & Submerged  & 1 & 10 & 2 & 22.92 $\pm$ 9.10 & 88.99 $\pm$ 12.58 & 5.14 $\pm$ 0.30 & 112.34 $\pm$ 9.25 & 1.20 $\pm$ 0.03 & 164.47 $\pm$ 25.46 & 47.78 $\pm$ 5.16 & 2.47 $\pm$ 1.49 & 0.20 $\pm$ 0.11 & 11.92 $\pm$ 1.18 & 0.0067 $\pm$ 0.0034 \\
Pregnant & Fresh & Submerged  & 0.5 & 0 & 3 & 24.85 $\pm$ 2.16 & 282.34 $\pm$ 12.33 & 6.15 $\pm$ 0.35 & 59.93 $\pm$ 41.39 & 2.02 $\pm$ 0.79 & 55.82 $\pm$ 21.46 & 55.96 $\pm$ 52.90 & 0.40 $\pm$ 0.09 & 0.18 $\pm$ 0.12 & 2.72 $\pm$ 1.07 & -- \\
Pregnant & Fresh & Submerged  & 0.5 & 2 & 3 & 19.29 $\pm$ 1.66 & 295.88 $\pm$ 9.26 & 6.39 $\pm$ 0.34 & 111.99 $\pm$ 48.38 & 1.59 $\pm$ 0.28 & 168.85 $\pm$ 75.57 & 57.84 $\pm$ 24.64 & 1.11 $\pm$ 0.72 & 0.27 $\pm$ 0.04 & 3.94 $\pm$ 2.27 & 0.0009 $\pm$ 0.0007 \\
Pregnant & Fresh & Submerged  & 0.5 & 5 & 3 & 34.00 $\pm$ 5.31 & 305.20 $\pm$ 19.27 & 5.98 $\pm$ 0.31 & 42.19 $\pm$ 12.20 & 1.75 $\pm$ 0.93 & 58.31 $\pm$ 22.01 & 30.49 $\pm$ 19.33 & 0.74 $\pm$ 0.55 & 0.21 $\pm$ 0.14 & 3.38 $\pm$ 0.78 & 0.0019 $\pm$ 0.0012 \\
Pregnant & Fresh & Submerged  & 0.5 & 10 & 3 & 24.58 $\pm$ 17.38 & 303.01 $\pm$ 2.70 & 6.02 $\pm$ 0.58 & 93.74 $\pm$ 53.96 & 1.90 $\pm$ 0.40 & 118.98 $\pm$ 81.93 & 55.03 $\pm$ 18.50 & 1.45 $\pm$ 1.66 & 0.43 $\pm$ 0.21 & 2.97 $\pm$ 2.43 & 0.0059 $\pm$ 0.0040 \\
Pregnant & Fresh & Submerged  & 0.5 & 15 & 3 & 24.12 $\pm$ 3.57 & 302.76 $\pm$ 18.72 & 5.45 $\pm$ 0.46 & 42.48 $\pm$ 25.20 & 2.29 $\pm$ 0.62 & 49.42 $\pm$ 33.05 & 42.72 $\pm$ 40.88 & 0.77 $\pm$ 0.65 & 0.35 $\pm$ 0.19 & 2.09 $\pm$ 1.27 & 0.0056 $\pm$ 0.0017 \\
Pregnant & Fresh & Submerged  & 1 & 0 & 2 & 25.66 $\pm$ 6.52 & 325.69 $\pm$ 16.47 & 6.85 $\pm$ 0.37 & 63.75 $\pm$ 25.31 & 1.23 $\pm$ 0.08 & 85.98 $\pm$ 22.35 & 32.35 $\pm$ 7.18 & 2.74 $\pm$ 2.82 & 0.24 $\pm$ 0.13 & 9.70 $\pm$ 6.96 & -- \\
Pregnant & Fresh & Submerged  & 1 & 10 & 3 & 19.90 $\pm$ 6.80 & 309.97 $\pm$ 27.63 & 5.93 $\pm$ 0.23 & 92.76 $\pm$ 36.23 & 1.48 $\pm$ 0.58 & 126.90 $\pm$ 35.64 & 53.98 $\pm$ 33.62 & 4.25 $\pm$ 5.78 & 0.23 $\pm$ 0.10 & 17.45 $\pm$ 23.06 & 0.0064 $\pm$ 0.0062 \\
Pregnant & Frozen & Submerged  & 1 & 5 & 2 & 31.25 $\pm$ 18.83 & 342.39 $\pm$ 10.42 & 6.35 $\pm$ 0.16 & 65.05 $\pm$ 53.87 & 1.22 $\pm$ 0.40 & 82.69 $\pm$ 38.19 & 39.90 $\pm$ 43.42 & 2.59 $\pm$ 3.02 & 0.15 $\pm$ 0.01 & 16.68 $\pm$ 19.09 & 0.0020 $\pm$ 0.0004 \\
Pregnant & Frozen & Submerged  & 1 & 10 & 2 & 37.19 $\pm$ 10.42 & 342.39 $\pm$ 10.42 & 7.80 $\pm$ 0.02 & 51.63 $\pm$ 3.08 & 1.73 $\pm$ 0.99 & 71.09 $\pm$ 31.96 & 35.14 $\pm$ 12.64 & 6.64 $\pm$ 7.85 & 0.33 $\pm$ 0.08 & 18.16 $\pm$ 19.92 & 0.0038 $\pm$ 0.0033 \\
Pregnant & Frozen & Submerged  & 1 & 15 & 2 & 24.56 $\pm$ 11.38 & 376.08 $\pm$ 0.00 & 7.01 $\pm$ 0.99 & 43.62 $\pm$ 11.03 & 1.90 $\pm$ 0.12 & 54.42 $\pm$ 4.94 & 33.64 $\pm$ 9.59 & 4.35 $\pm$ 2.73 & 0.54 $\pm$ 0.29 & 7.88 $\pm$ 1.02 & 0.0048 $\pm$ 0.0007 \\
\end{tabular}
}
\end{table}
\end{landscape}

\FloatBarrier

\subsection{Raw data curves for all tests}

\begin{figure}[htbp!]
    \centering
    \includegraphics[width=1\textwidth]{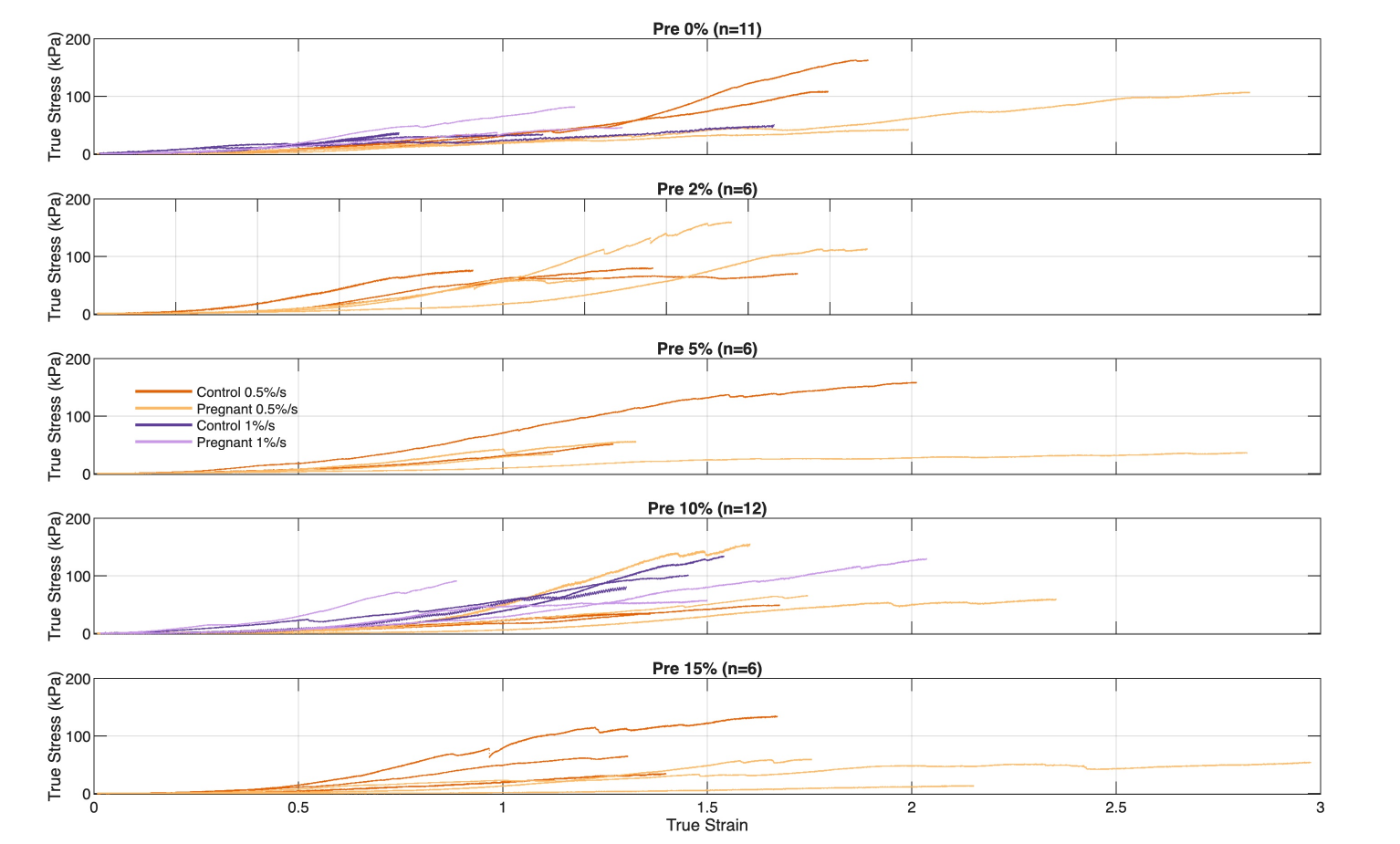}
    \caption{True stress vs. strain plots for fresh pelvic floor samples from pregnant and control (non-pregnant) rats with A. no preconditioning, 10 cycles of B. 2\%, C. 5\%, D. 10\%, and E. 15\% engineering strain for preconditioning prior to failure testing. 
    Data is plotted up to peak stress.}
    \label{fig:FigS_PreconditionComparisonSSCurves}
\end{figure}

\begin{figure}[htbp!]
    \centering
    \includegraphics[width=1\textwidth]{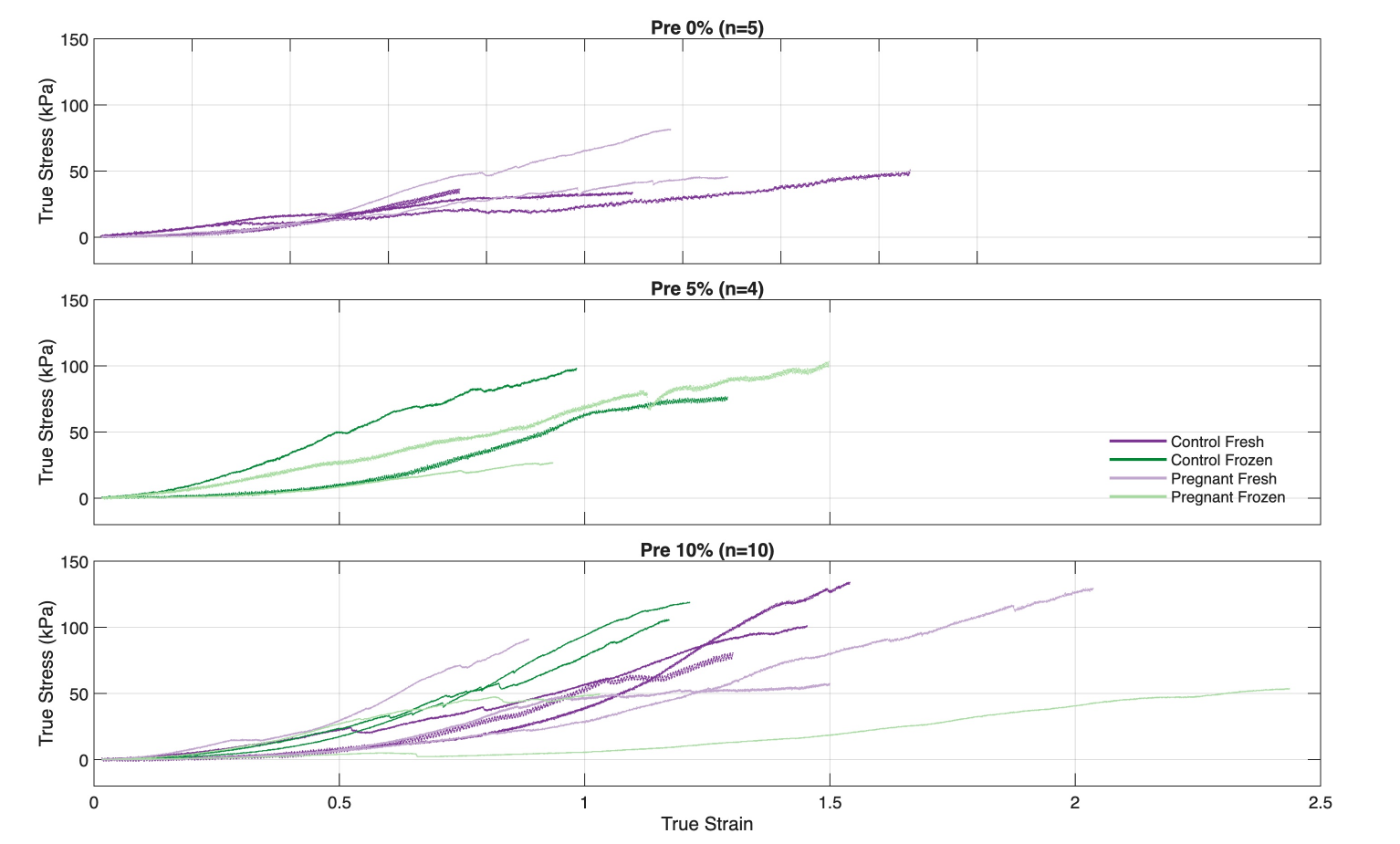}
    \caption{True stress vs. strain plots for fresh pelvic floor samples from pregnant and control (non-pregnant) rats with A. no preconditioning, 10 cycles of B. 5\% and C. 10\% engineering strain for preconditioning prior to failure testing. 
    Data is plotted up to peak stress.}
    \label{fig:FigS_FreezingComparisonSSCurves}
\end{figure}

\Cref{fig:FigS_PreconditionComparisonSSCurves} and \Cref{fig:FigS_FreezingComparisonSSCurves} show the raw data curves for all experimental conditions used in this study, and directly corresponds with the box plots in \Cref{fig:FigR_PreconditionComparisonBoxplotPanel}--\Cref{fig:FigR_FreezingComparisonBoxplotPanel}. Individual curves are plotted to bulk failure for each sample used for comparisons in the boxplots. 

\FloatBarrier

\subsection{Cyclic peak to peak stresses}

\begin{figure}[htbp!]
    \centering
    \includegraphics[width=1\textwidth]{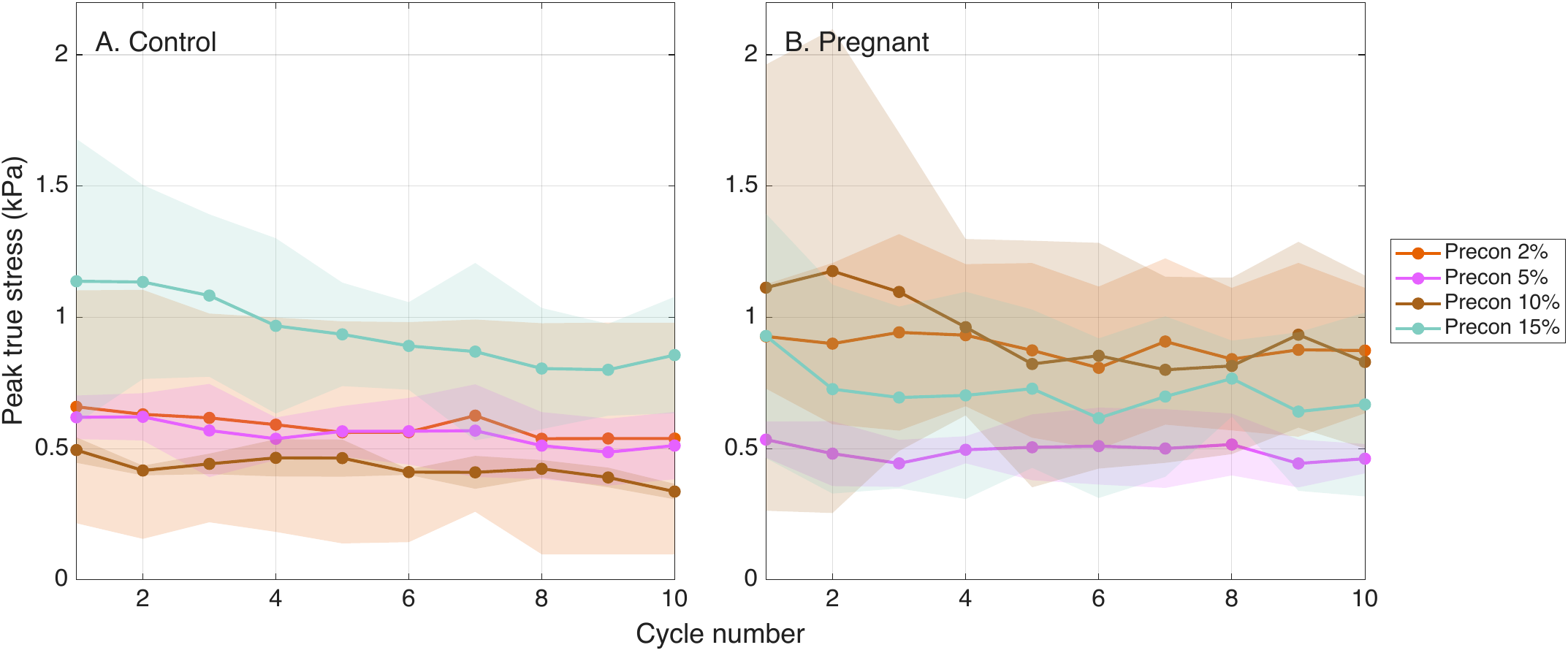}
    \caption{Peak true stress from cycle to cycle for the 10 cycles of preconditioning done at each rate for A. control, and B. pregnant tissue, with shaded regions showing standard deviation of each condition.}
    \label{fig:FigS_PeakCyclic}
\end{figure}

\begin{figure}[htbp!]
    \centering
    \includegraphics[width=1\textwidth]{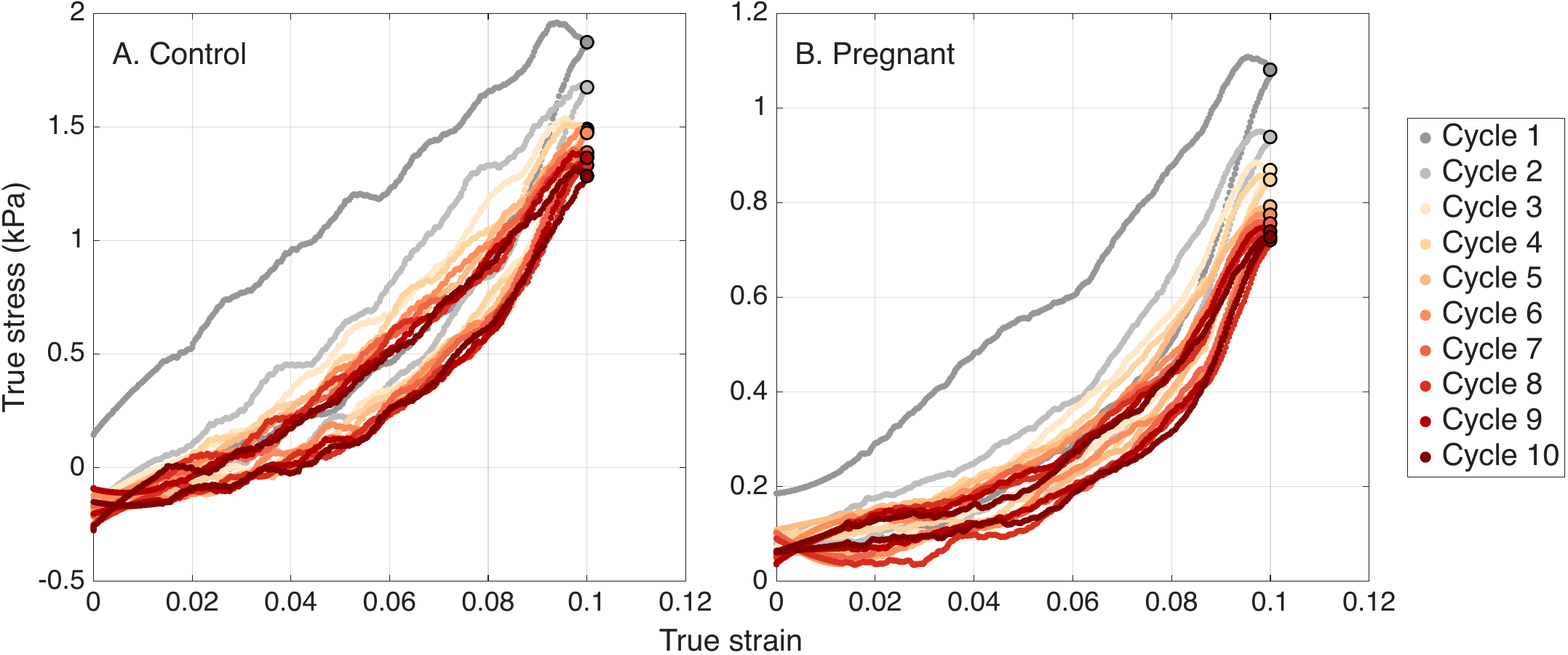}
    \caption{Preconditioning stress versus strain curves showing smoothed cycle to cycle material response equilibration after 10 cycles for A. Control and B. Pregnant representative curves.}
    \label{fig:FigS_PreconCyclic}
\end{figure}

\Cref{fig:FigS_PeakCyclic} shows the peak stress during each preconditioning cycle for samples tested at 2\%, 5\%, 10\%, and 15\% strain amplitude, and \Cref{,fig:FigS_PreconCyclic} shows the smoothed representative control and pregnant preconditioning stress-strain curves for each cycle.  Notice that the results approximately equilibrate in around 5-7 cycles for both pregnant and control samples, regardless of the strain amplitude used for the preconditioning. 

\FloatBarrier

\subsection{Representative stress-strain curve with area uncertainty taken into account}

\begin{figure}[htbp!]
    \centering
    \includegraphics[width=1\textwidth]{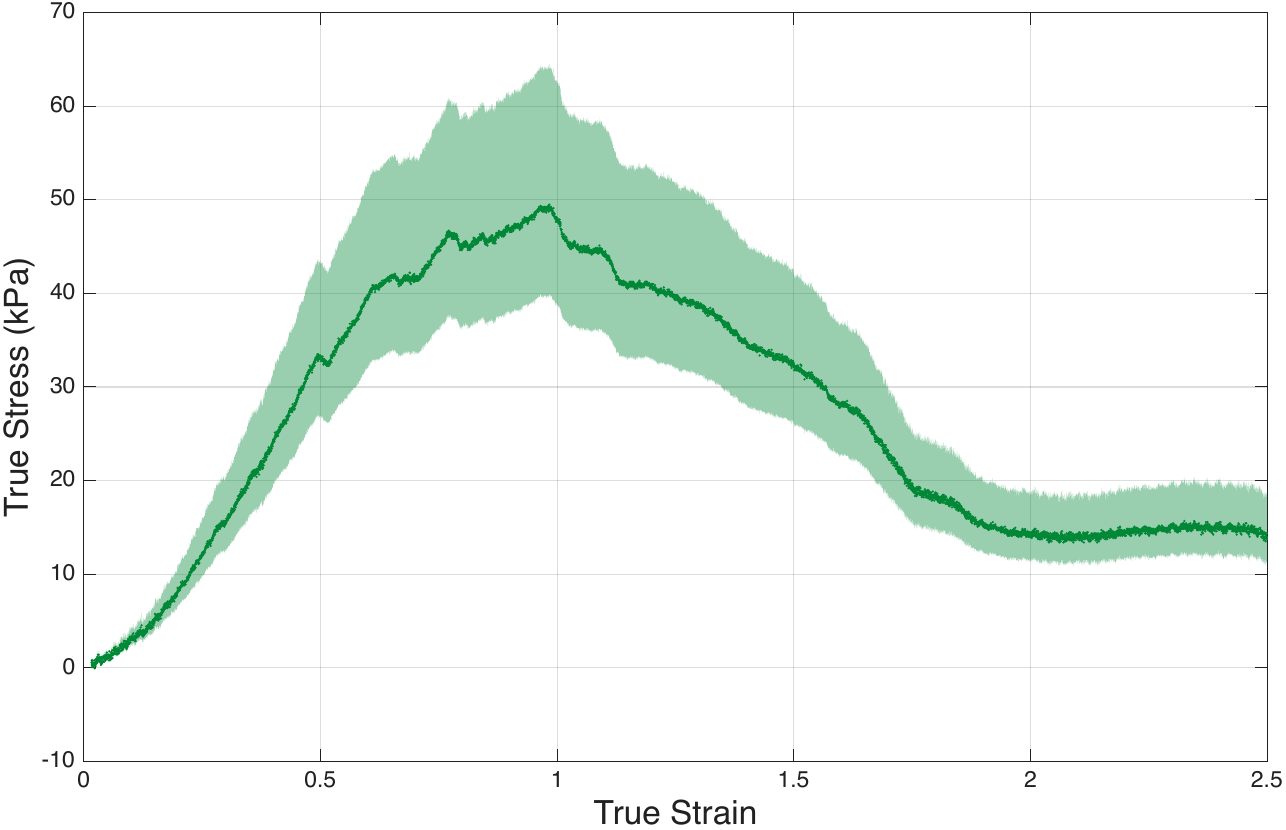}
    \caption{Representative curve of control sample loaded at 0.5\% strain per second with no preconditioning and previously frozen showing resultant variation in stress data (shaded region) based on variability of measured sample gauge width and thickness.}
    \label{fig:FigS_RepresentativeCurveWithUncertainty}
\end{figure}

\Cref{fig:FigS_RepresentativeCurveWithUncertainty} shows a representative raw stress-strain curve for a single sample that takes into account the variability in the cross-sectional area measurements. This high degree of uncertainty in the area propagates to high uncertainty in the stress and suggest localized areas may be needed to minimize this variability.  

\FloatBarrier

\subsection{Representative Curve with localized fractures}

\begin{figure}[htbp!]
    \centering
    \includegraphics[width=1\textwidth]{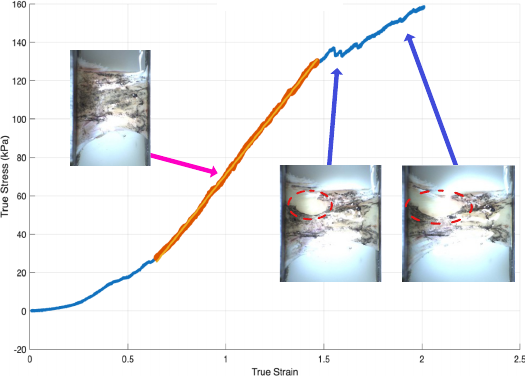}
    \caption{Representative curve of control sample loaded at 0.5\% strain per second with 10 cycles of 2\% strain-amplitude preconditioning loaded to failure, with top-down images of the sample during testing showing the linear region loading, a post-stepping mini fracture (with an example of that mini fracture circled with a red dotted line) before bulk fracture, and continued expansion of this (and other) fracture as loading increases up to bulk failure.}
    \label{fig:FigS_MiniFractures}
\end{figure}

\Cref{fig:FigS_MiniFractures} shows a representative sample stress-strain curve for a sample with overlayed linear region stiffness, along with corresponding images of the same sample showing mini fractures that approximately coincide with dips/stepping in the stress-strain curve. 

\FloatBarrier

\subsection{Methods Comparison to literature}
\begin{landscape}
\small
\renewcommand{\arraystretch}{1.3}

\begin{longtable}{p{3.2cm} p{5.0cm} p{5.0cm} p{5.0cm}}

\caption{Comparison of experimental and analysis methods used in the present rat LAM study and human cadaveric LAM literature datasets.}
\label{tab:TableS_MethodsComparison_Literature} \\

\toprule
\textbf{Category} &
\textbf{Present study} &
\textbf{Nagle et al. 2014} &
\textbf{Routzong et al. 2026} \\
\midrule
\endfirsthead

\caption[]{Comparison of experimental and analysis methods used in the present rat LAM study and human cadaveric LAM literature datasets. Continued.} \\

\toprule
\textbf{Category} &
\textbf{Present study} &
\textbf{Nagle et al. 2014} &
\textbf{Routzong et al. 2026} \\
\midrule
\endhead

\midrule
\multicolumn{4}{r}{Continued on next page} \\
\endfoot

\bottomrule
\endlastfoot

Tissue tested &
Rat LAM complex. &
Post-menopausal human LA segments fresh-frozen. &
Human LAM isolated from en bloc complexes fresh-frozen. \\

Storage / thawing &
Stored at $4^\circ$C (fresh) or thawed overnight at $4^\circ$C (frozen). &
Cadavers thawed for three days prior to dissection/testing. &
Torsos/complexes thawed over 2--3 days prior to dissection/testing. \\

Hydration / temperature &
Tested submerged in 1X PBS at $37^\circ$C. &
Soaked in saline, wrapped in gauze; tested in air at room temperature within 4 hours. &
Hydrated prior to testing; tested in air at room temperature (to avoid DIC interference). \\

Area measurement &
Width and thickness measured from specimen images (ImageJ); averaged for cross-sectional area. &
Width and thickness measured via digital calipers at 3 locations and averaged. &
Width measured via digital calipers; thickness via laser displacement sensor at 3 locations. \\

Preconditioning &
10 cycles (triangular-wave) at 0\%, 2\%, 5\%, 10\%, or 15\% engineering strain amplitude, then 15-min rest. &
10 cycles (sinusoidal) with 2.54 mm amplitude ($\sim$10\% expected gauge length). &
No preconditioning protocol described. \\

Loading protocol &
Uniaxial loading to failure (0.5\% or 1\% engineering strain/sec). &
Monotonic uniaxial extension to failure (1 mm/s). &
Uniaxial tensile testing to specimen/clamp failure (5 mm/min). \\

Processing / smoothing &
Automated MATLAB pipeline; Savitzky--Golay smoothing for transition-region detection. &
11-point moving average smoothing; used for constitutive model fitting/yield analysis. &
Trimmed to remove post-failure/clamp-failure; custom MATLAB pipeline for DIC noise reduction. \\

\end{longtable}
\end{landscape}


\bibliographystyle{elsarticle-harv}
\bibliography{references}

\end{document}